\documentclass[a4paper,11pt]{article}
\pdfoutput=1 

\usepackage{jheppub} 
\usepackage{physics}
\usepackage{xcolor}
\usepackage[T1]{fontenc} 
\usepackage{graphicx}
\usepackage{subcaption}
\usepackage[normalem]{ulem}
\usepackage{epstopdf}

\title{\boldmath Complexity in One- and Two-Qubit Systems}

\author[a,b]{Reginald J. Caginalp}
\author[c]{and Samuel Leutheusser}

\affiliation[a]{Berkeley Center for Theoretical Physics, University of California, Berkeley, CA 94720, USA}
\affiliation[b]{Physics Division, Lawrence Berkeley National Laboratory, Berkeley, CA 94720, USA}
\affiliation[c]{Center for Theoretical Physics, Massachusetts Institute of Technology, Cambridge, MA 02139, USA}

\emailAdd{caginalp@berkeley.edu}
\emailAdd{sawl@mit.edu}

\preprint{MIT-CTP/5245}

\abstract{We numerically analyze the complexity of unitary time-evolution and precursor operators in one- and two-qubit systems using the framework of Nielsen complexity geometry. We find that, as expected, the complexities of unitary time evolution operators grow linearly with time, at least initially, for both the one- and two-qubit cases. The precursor operators display switchback-effect-like behavior provided we choose our cost factors so that the resulting complexity geometry is negatively curved.}

\begin{document} 
\maketitle
\flushbottom

\section{Introduction}

Recent work has shown that information plays a fundamental role in gravity, holography and the structure of spacetime. The crucial role of information is often most easily studied in the context of the Anti-de Sitter (AdS)/Conformal Field Theory (CFT) correspondence, which posits an equivalence between quantum gravity in $(d+1)-$dimensional AdS space and a CFT in $d$ dimensions~\cite{Maldacena, Gubser, Witten}. An explicit example is the Ryu-Takayanagi (RT) formula~\cite{RT,HRT}, which states that the entanglement entropy of a CFT subregion is equivalent to the area of a minimal surface in AdS anchored on the entangling surface of the CFT subregion. This information-theoretic entry in the AdS/CFT `dictionary' has led to many important results in understanding holographic duality including, for example, entanglement wedge reconstruction~\cite{ADH,DHW}. Such advances have motivated a push to understand how other information-theoretic probes are realized in a theory of quantum gravity.

A recent proposal along these lines regards the role that quantum complexity may play in holographic duality~\cite{CompVol, Brown, Brown2}. Given a state $\ket \Psi$ in a Hilbert space $\mathcal{H}$, the complexity $\mathcal C ( \ket \Psi )$ is the minimum number of ``simple'' gates that one must act on a ``simple'' reference state $\ket{ \Psi_0 }\in \mathcal{H}$ to obtain $\ket \Psi$. For example, in a system of qubits, ``simple''  gates might be chosen to be those that act on a small number of qubits, and a reference state might by the untangled state $\ket { 000 \cdots 0}$. A closely-related concept is the complexity of a unitary operator, which is the minimum number of ``simple'' gates we need to compose to obtain the desired unitary operator.  We discuss the approach to operator complexity taken in this paper, Nielsen complexity geometry, in more detail in Section~\ref{sec:comp_geom}.

\begin{figure}
\centering
 \includegraphics[width = 0.75\textwidth]{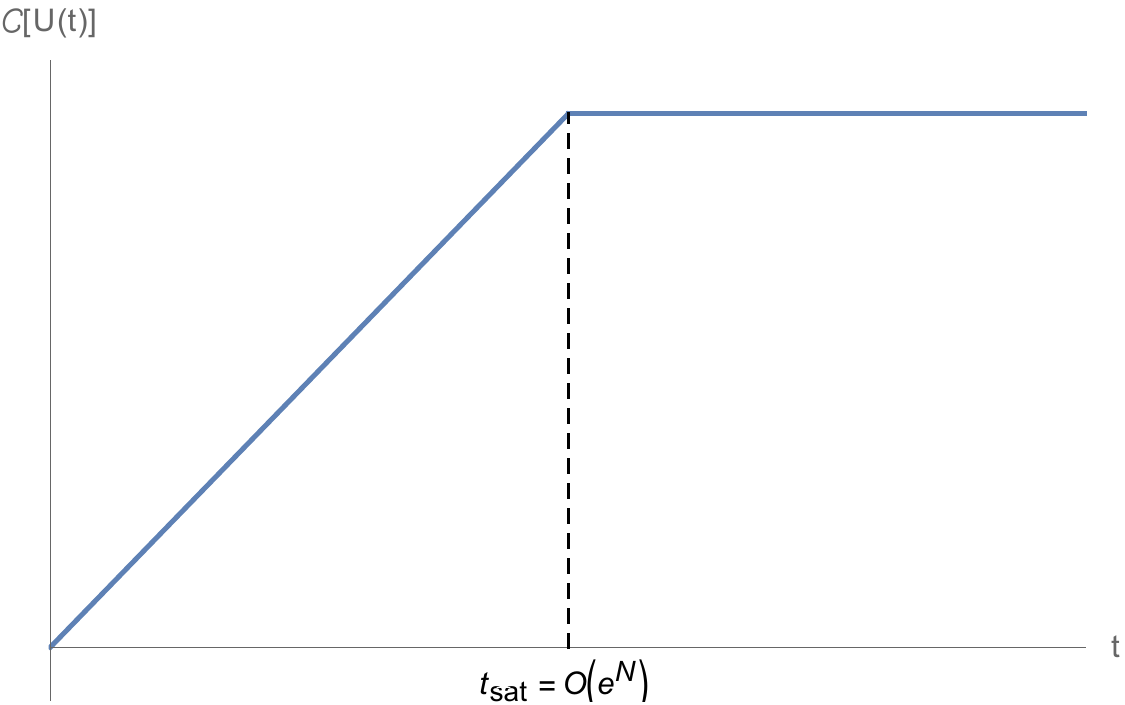}
  \caption{Schematic of complexity of a unitary time-evolution operator versus time. Initially, the complexity grows linearly with time, before saturating at some time that is exponential in the number of degrees of freedom $N$. At very late times, there will be Poincare recurrences (not shown here).}\label{fig:U_compl_schem}
\end{figure}

Although the precise values of complexity depend on details such as the tolerance, the precise set of ``simple'' gates chosen, and so on, complexities obey some simple qualitative features. For example, a system of a large number of qubits with a generic Hamiltonian $H$ has time-evolution unitary $U(t) = \exp(-iHt)$. In this scenario, one expects that the complexity of $U(t)$ will grow linearly with time, at least initially. However, at a certain point, the complexity will saturate and be approximately constant thereafter~\cite{LPiTP}. This was shown explicitly by~\cite{SYKComp}. See Figure~\ref{fig:U_compl_schem}.   Previous work by Brown and Susskind argued for a thermodynamics of complexity~\cite{2ndLaw}. In particular, they argued that the complexity of a quantum system with $N$ qubits is related to an entropy of a classical system with $2^N$ degrees of freedom. They argue for a ``second law of complexity,'' which says that any system with non-maximal complexity will be overwhelmingly likely to increase its complexity. Moreover, they argue that a system that does not have maximal complexity can use this as a resource (which the authors of~\cite{2ndLaw} dub ``uncomplexity'') for quantum computation.

\begin{figure}
\centering
 \includegraphics[width = 0.75\textwidth]{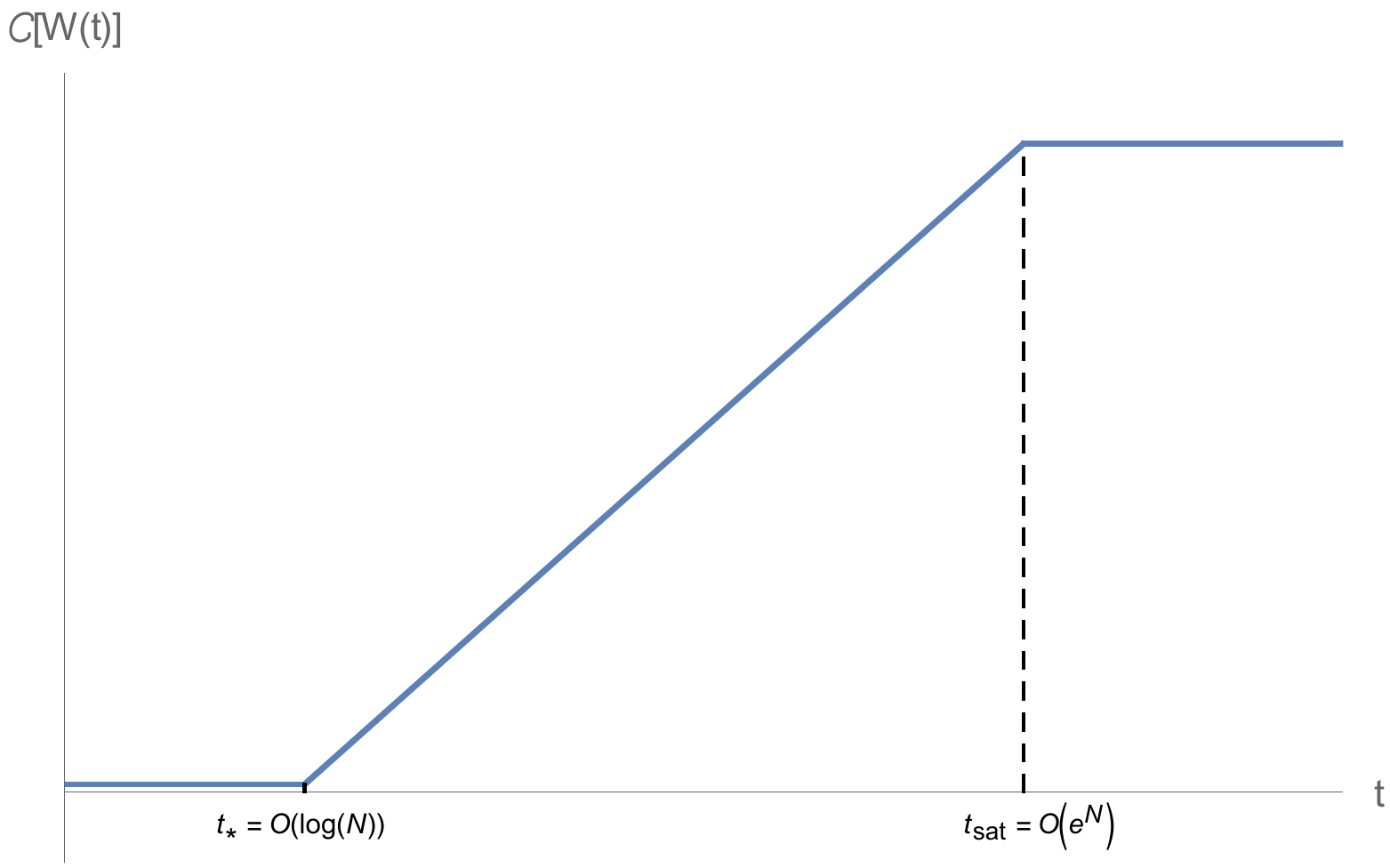}
  \caption{Schematic of complexity of a precursor operator versus time. Initially, the complexity is very small, up until the scrambling time $t_*$, at which point it begins to grow linearly with time, before saturating at some time that is exponential in the number of degrees of freedom $N$. At very late times, there will be Poincare recurrences (not shown here).}\label{fig:W_compl_schem}
\end{figure}

Another interesting feature of complexity of these systems is the so-called switchback effect~\cite{Switchback,QCNC}. This involves precursor operators, which are the time-evolved ``simple'' operators. As an example, for a simple `initial' operator $W_0$ (on a qubit system $W_0$ could be $X \otimes I \otimes \cdots \otimes I$), the precursor operator $W(t)$ is 
$$W(t) = e^{-iHt} W_0 e^{iHt}.$$
Initially, the operator $W(t)$ is not very complex, since it still is mostly along the ``simple'' direction, $W_0$. However, after a time of order a scrambling time, the complexity begins to grow linearly~\cite{Switchback, QCNC}. This delay in complexity growth on the order of a scrambling time is the switchback-effect. It can be intuitively understood  as the delay for $W(t)$ to be begin to be supported on ``complicated'' operators since at small times $W(t)$ will consist primarily of ``simple'' gates which do not contribute to its complexity. In particular, consider the small-$t$ expansion of the above precursor. By the Baker-Campbell-Hausdorff formula, there will be nested commutators of $W_0$ and $H$. $H$ is a sum of terms that act on a small number of qubits. As $t$ increases, the terms with larger numbers of nested commutators increase. These terms with a higher number of nested operators will generate terms with support on a larger number of qubits, and will thus have high complexity. As these high-complexity terms are multiplied by higher powers of $t$, it takes a larger amount of time for them to become significant. Once $W(t)$ has support on a large number of degrees of freedom, its complexity will of course be large. See Figure~\ref{fig:W_compl_schem}. A key element of the switchback effect in large $N$ systems is the negative curvature of complexity geometry~\cite{Switchback,QCNC}, which we explain in greater detail in later sections.

 There has been much recent work exploring the role complexity may play in holographic systems. In a two-sided AdS black hole, it was observed that the entanglement entropy of one side quickly saturates~\cite{Hartman}. However, the volume of the maximum slice of the black hole keeps growing forever, at least classically. The AdS/CFT duality then asserts that this volume should be dual to some CFT quantity. The essence of the holographic complexity proposals~\cite{CompVol, Brown, Brown2} is that the dual of this quantity in the CFT side is the computational complexity.

The two main proposals are the complexity-volume (CV) proposal~\cite{CompVol} and the complexity-action (CA) proposal~\cite{Brown,Brown2}. Consider a state $\ket \Psi $ in a holographic CFT with a semi-classical dual. The complexity-volume proposal says that the complexity of the state $\ket \Psi$ is equal to the volume of the maximal volume slice (i.e., co-dimension 1 surface) through the bulk. The complexity-action proposal says that the complexity of $\ket \Psi$ is the action of the Wheeler-de Witt patch, which is defined as the domain of dependence of the maximum volume slice.

Although these proposals are somewhat speculative, they pass some non-trivial checks. For example, some recent work has investigated complexity in simple quantum field theories such as the free scalar field~\cite{Chapman17,Jefferson}. These are quite different from the types of strongly-coupled, large-$N$ field theories that are expected to have semi-classical holographic duals, but there are some intriguing similarities between these. In particular, in the analysis of the states in free quantum field theories, there are ambiguities in the reference state. It is tempting to identify these with the ambiguities in the bulk gravitational action~\cite{Lehner}. Moreover, there is a logarithmic divergence in the UV cutoff in the field theory calculation, which is similar to the logarithmic divergence in the near-boundary cutoff in the complexity-action bulk calculation. 

Importantly, these holographic complexity proposals also reproduce the switchback effect~\cite{Switchback, QCNC}, with the standard scrambling time for black holes, $t_* = \frac{\beta}{2 \pi} \log N^2$, where $\beta$ is the inverse temperature of the black hole, and $N$ is the rank of the gauge group on the CFT side of the duality~\cite{Shenker,Stanford}. Namely, the complexity of a precursor is low up until the scrambling time $t_*$, at which point it begins to grow linearly.

\begin{figure}
\centering
 \includegraphics[width = 0.75\textwidth]{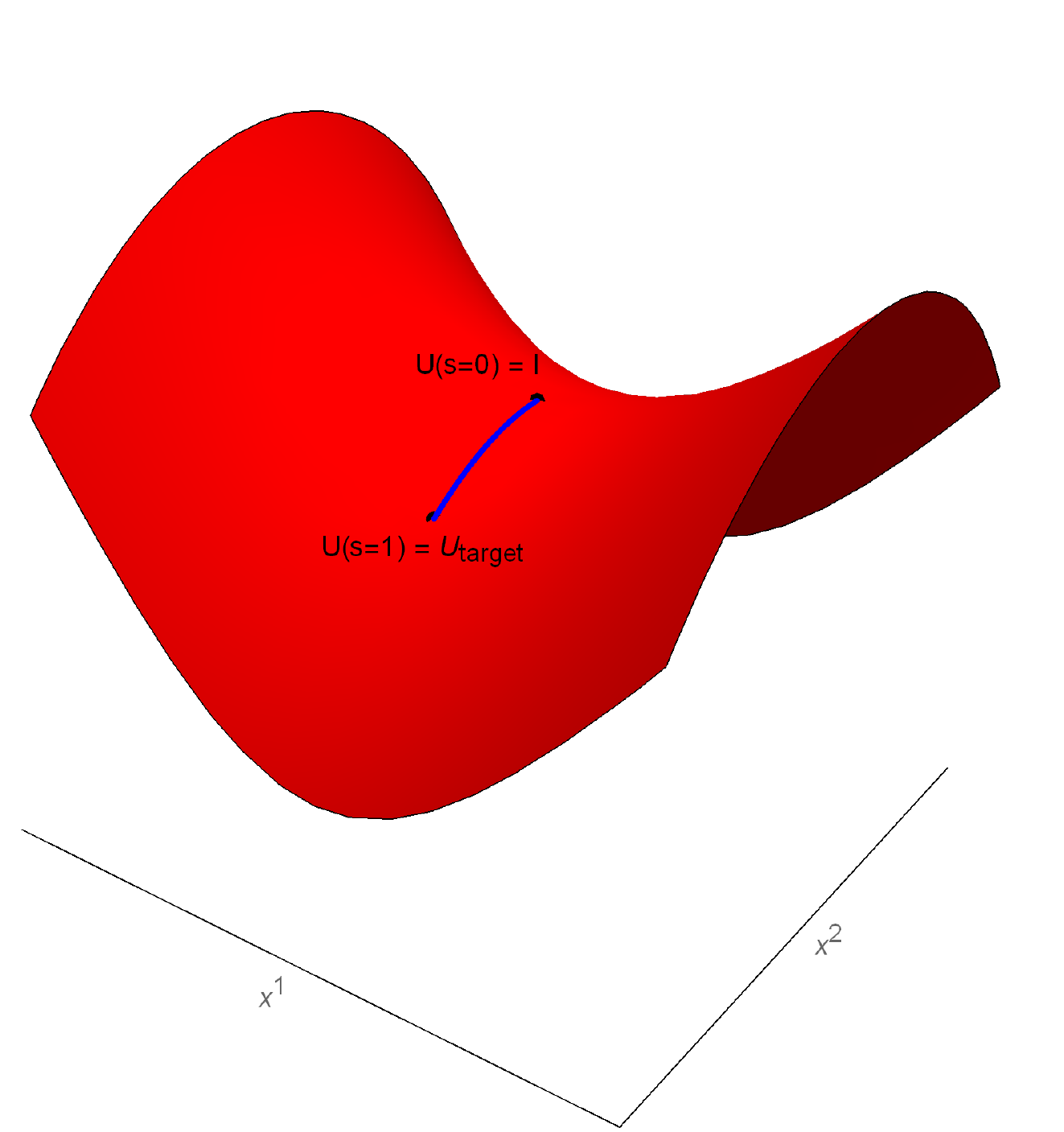}
  \caption{Schematic of complexity geometry. The intrinsic curvature, described by the complexity metric, of the unitary manifold (red surface) covered by coordinate patch $(x^1,x^2)$ is visualized as extrinsic curvature in an embedding space. The blue line is the minimum-length path between the identity operator $I$ and the target unitary, $U_{\text{target}}$.}\label{fig:Space_schematic}
\end{figure}

The purpose of this paper is to help bridge the gap between these qualitative expectations of the CV and CA proposals, and notions of complexity in the boundary theory of holographic systems. Specifically, we will calculate the complexity of various operators in one- and two-qubit systems, using the framework of Nielsen complexity geometry, and analyze their behavior in time. This will allow us to quantitatively calculate complexity and compare with the qualitative intuition above. In this approach, we assign a metric to the space of unitaries, making some directions have a much larger cost than others. In a large-$N$ system, we may assign gates that act on more than two degrees of freedom have a very large cost. The complexity of a given unitary $U$ is then given by the length of the minimal-length path (i.e., geodesic) between the identity operator $I$ and the target unitary $U$. See Figure~\ref{fig:Space_schematic}

For the operators we consider in this paper, we will find the following features: the complexity of unitary time evolution operators grows linearly with time, at least initially, as expected. For choices of complexity geometry that are negatively curved, the precursors show the qualitative behavior described above. The initial rate of growth is slow and then it begins quicken, entering a regime of linear growth of complexity in time; however, the distinction between the two regimes is not as pronounced as it is in the large-$N$ case.

Our calculations provide support for the aforementioned conjectures in a concrete setting with explicit computations. In particular, we provide a first numerical study of precursor complexity in small systems. Previous work analyzed the geodesics in complexity space using the Euler-Arnold formalism~\cite{BindComp, SYKComp}.  The paper~\cite{BindComp} analyzed the ``binding complexity'' of various qubit states that are analogous to multi-boundary wormholes in AdS/CFT. These calculations showed that the binding complexities behave similarly to the expectations from the holographic complexity proposals. The work in~\cite{SYKComp} considered the complexity of unitary time-evolution operators for one-qubit systems, as well as an analysis of the SYK model.

We begin with a brief review of Nielsen complexity geometry. We then investigate complexity in one-qubit geometries, considering a geometry where the $Z$-direction is much harder than the $X$ and $Y$-directions. We find that the complexity of the unitary time-evolution operator grows linearly with time for early times. The complexity of precursors initially grows slowly, before then growing linearly, mimicking the behavior of large-$N$ systems discussed above. We discuss how precursors are sensitive to the degree of the anisotropy in unitary space, while the time-evolution operators are not. We then consider two-qubit systems. As expected, at early times, complexity of the time-evolution unitary grows linearly with time. If we choose one-qubit gates to be easier and a Hamiltonian consisting of single qubit terms, then the complexity of a precursor is constant with time. However, if we take the two-qubit gates to be easier and a Hamiltonian consisting of two-qubit terms, we find that the complexity of precursors grows slowly at first, and then begins to grow linearly. Although this latter choice is unusual from a laboratory perspective, it gives a better model of the switchback effect. We use units where $\hbar =1$ throughout.

\section{Nielsen Complexity Geometry} 
\label{sec:comp_geom}

We briefly review Nielsen complexity geometry which we will be using throughout~\cite{Nielsen1,Nielsen2,Nielsen3,Nielsen4,Nielsen5}. Suppose that we have a system of $N$ qubits, and a general unitary $U$ in the group of unitaries that act on the $N$ qubits, $SU(2^N)$. Any such unitary, $U$, can be parameterized by $4^N-1$ parameters $x_1, x_2, \ldots, x_{4^N-1}$ indicating the extent of that unitary along each generator of $SU(2^N)$. A general metric on this space of unitaries is given by 
\begin{equation}
ds^2 = \text{Tr} (i dU U^\dagger T_I) \mathcal{I}_{IJ} \text{Tr} (i dU U^\dagger T_J),
\label{eq:metric_def}
\end{equation}
where $T_I$ is a generator of $SU(2^N)$ (which later we will take to be a tensor product of Pauli matrices), and $ \mathcal{I}_{IJ}$ is the cost factor. We will take $ \mathcal{I}_{IJ}$ to be diagonal; hence, $ \mathcal{I}_{IJ}$ characterizes the `hardness' of applying the generator $T_I$. With the parametrization by coordinates, $x^I$, the corresponding group element $U \in SU(2^N)$ is given by 
$$U(x^I) = \exp( i x^I T_I ).$$
To find the complexity of a unitary, $U_{target}$, we first need to solve the geodesic equation
$$\frac{d^2 x^I}{ds^2} = - \Gamma^I_{JK} \frac{d x^J}{ds} \frac{d x^K}{ds}$$
subject to the boundary conditions 
$$U(x^I(s=0)) = I,~ U(x^I(s=1)) = U_{target},$$
where the $\Gamma^I_{JK}$ are the usual Christoffel symbols for the Nielsen complexity metric~\ref{eq:metric_def}. The complexity of our target unitary will then by given by the length of the minimal geodesic,
$$\mathcal{C}[U_{target}] = \int_0^1 \sqrt{g_{IJ} \frac{d x^I}{ds} \frac{d x^J}{ds}} ds,$$
where $g_{IJ}$ is the Nielsen complexity metric.

In the later sections, we generate random Hamiltonians satisfying certain conditions, and then compute the complexity of the time-evolution unitary and of precursors with Nielsen metric such that ``easy'' gates have a lower cost factor than ``hard'' gates. We will always take the Hamiltonian to be a sum of ``easy'' terms. 

In our subsequent discussions, we will often have a target unitary and need to calculate its coordinates. Orthogonality of the $SU(2^N)$ generators:
$$\text{Tr} ( T_I T_J)  = 2^N \delta_{IJ},$$
allows us to compute the coordinates of a unitary $U_{target}$ via
\begin{equation}
i x^I = 2^{-N} \text{Tr}(\log(U_{target}) T_I).
\label{eq:coord}
\end{equation}
We can then numerically solve the geodesic equation, subject to the boundary conditions that the geodesic starts at the identity ($x^I=0$) and ends at the coordinates of the target unitary. 
We numerically calculate the values of the metric and Christoffel symbols. We use the Matlab function \texttt{bvp4c} to solve the boundary value problem given by the system of ODE's resulting from the geodesic equation. The boundary conditions are $x^I(s=0)=0$, corresponding to the identity, and $x^I(s=1)$ being the coordinates of target unitary, given by~\ref{eq:coord}. The function \texttt{bvp4c} uses a collocation method to solve the system of ODE's with boundary conditions. The function subdivides the interval [0,1] into subintervals, and discretizing the system of ODE's and boundary conditions, which turns it into an algebraic system of equations for values of the solution at the mesh points. It then numerically solves these algebraic systems.

\section{One Qubit System}

In this section, we consider complexity in a one-qubit system. Specifically, we consider the setup explored in \cite{SingleQubit}, where the $X$- and $Y$-directions are considered ``easy'' and the $Z$-direction is considered ``hard''. The $Z$-direction is taken to be 10 times harder than the $X$- and $Y$-directions.

\begin{figure}
\centering
 \includegraphics[width = 0.75\textwidth]{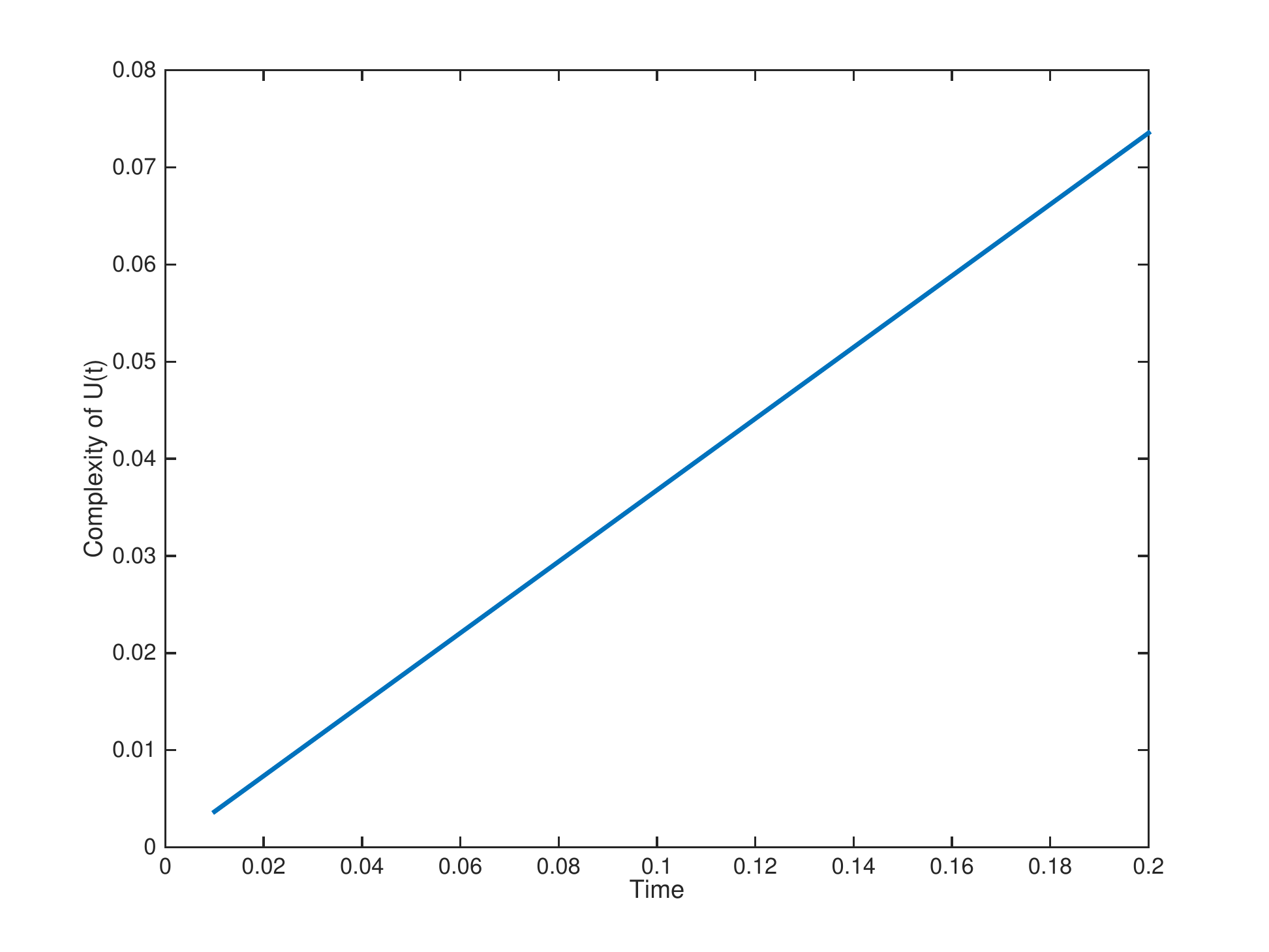}
  \caption{Complexity of a unitary time-evolution operator versus time for a one qubit system described in the main text. The complexity of this operator grows linearly with time, at least initially. The time axis is in units of seconds.}\label{fig:Unit_one_qubit}
\end{figure}

First, we take our Hamiltonian to be a sum of the easy terms, explicitly,
$$H = J_1 X + J_2 Y,$$ 
and consider the unitary time-evolution operator $U(t) = e^{ -i H t}$. We numerically solve the geodesic equations as described above for a sample Hamiltonian, and plot the complexity $\mathcal{C} [ U(t) ]$ as a function of time in Figure~\ref{fig:Unit_one_qubit}. We use the values $J_1 = 0.4387 s^{-1}, J_2 = 0.3816 s^{-1}$, and $\mathcal{I}_{XX} = \mathcal{I}_{YY}=0.1,~ \mathcal{I}_{ZZ} = 1$.  However, our answers for complexity will be equivalent for various classes of $J_i$. For the time-evolution operator, $U(t)$, for a fixed time $t$, the geodesic will lie in the isotropic subspace spanned by the $X$ and $Y$ directions. Hence, if we apply a rotation to $J_1, J_2$, the complexity will not change--the only thing that changes the complexity is an overall rescaling of $J_i$, which can absorbed into a rescaling of time $t$. A similar story applies for precursors, in that the complexity will be equivalent for various classes of choices of $J_i$ and $W_0$. See Appendix~\ref{app:A} for more details.

 Initially, before any recurrence time, we see the complexity of $U(t)$ grows linearly with time, exactly as expected. 
\begin{figure}
\centering
 \includegraphics[width = 0.75\textwidth]{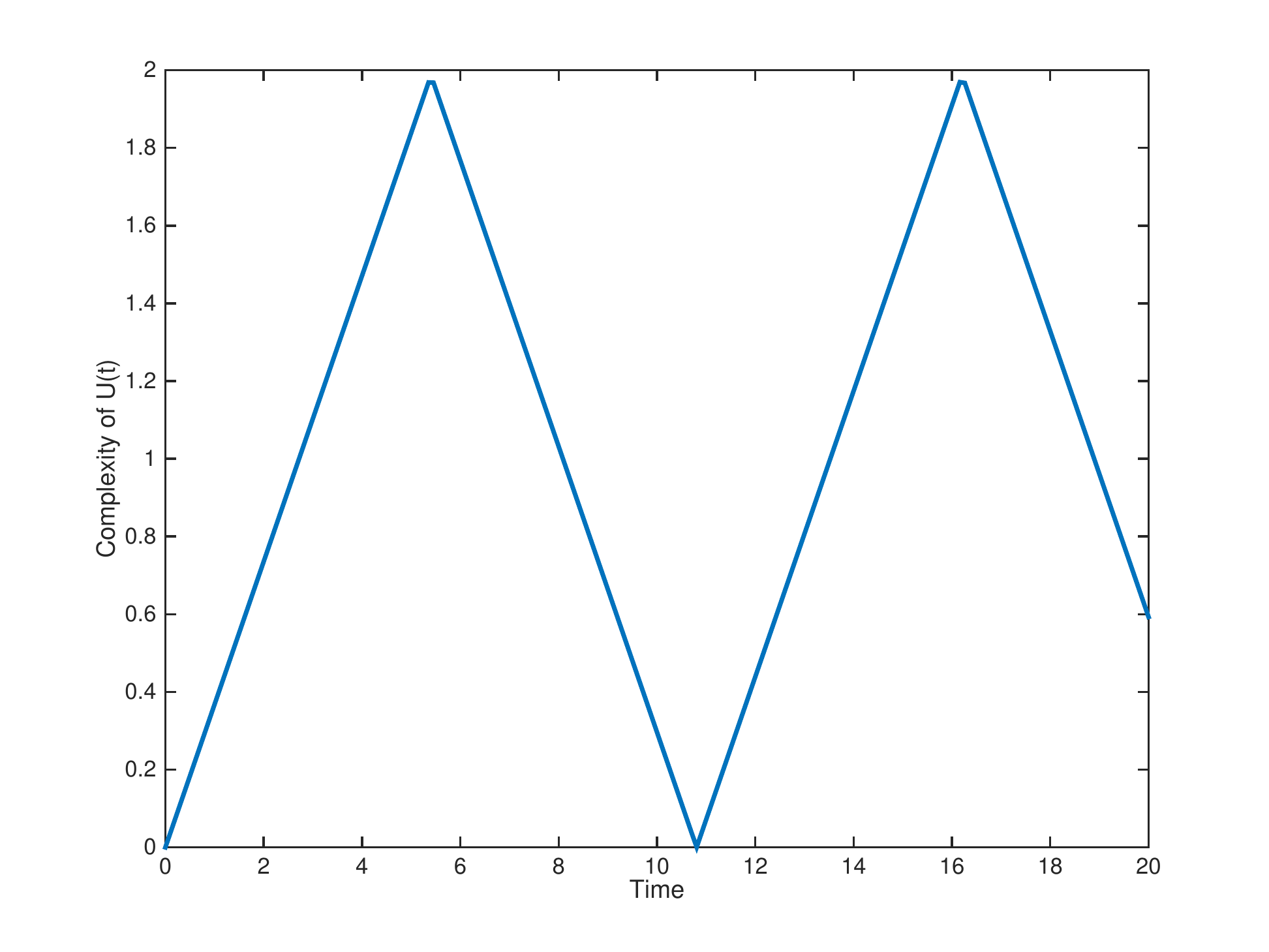}
  \caption{Complexity of a unitary time-evolution operator versus time for the one qubit system described in the main text. The complexity of this operator grows linearly with time, then linearly decreases to zero, at the Poincar\'e recurrence time. This process repeats itself indefinitely. The time axis is in units of seconds.}\label{fig:Unit_one_qubit_ext}
\end{figure}
However, as $t$ increases, there are topological obstructions or `conjugate points' and the length of the shortest path from $I$ to $U(t)$ stops increasing. To illustrate this point with a concrete example, we plot the complexity of $U(t)$ for a larger range of time in Figure~\ref{fig:Unit_one_qubit_ext}. The complexity initially grows linearly with time, until it reaches a maximum value, at time $t=\pi/\sqrt{J_1^2+J_2^2}$. For our Hamiltonian, this occurs at $t=5.4031 s.$ We discuss this in more detail in Appendix~\ref{app:A}. It then starts to decrease linearly to zero, at which point it begins to increase linearly again. This process repeats itself indefinitely. This agrees with the qualitative behavior of the complexity of a one-qubit unitary obtained in \cite{SYKComp}. These dips in the complexity are the one-qubit versions of Poincar\'e recurrences. As discussed in the introduction, in a system with $N$ degrees of freedom, we expect that the complexity will grow linearly up until a time that is $O(\exp(N))$. In our numerical work, we will only be concerned with regimes well before the appearance of topological obstructions or conjugate points.

\begin{figure}

\centering
\begin{subfigure}{.45\textwidth}

 \includegraphics[width = \textwidth]{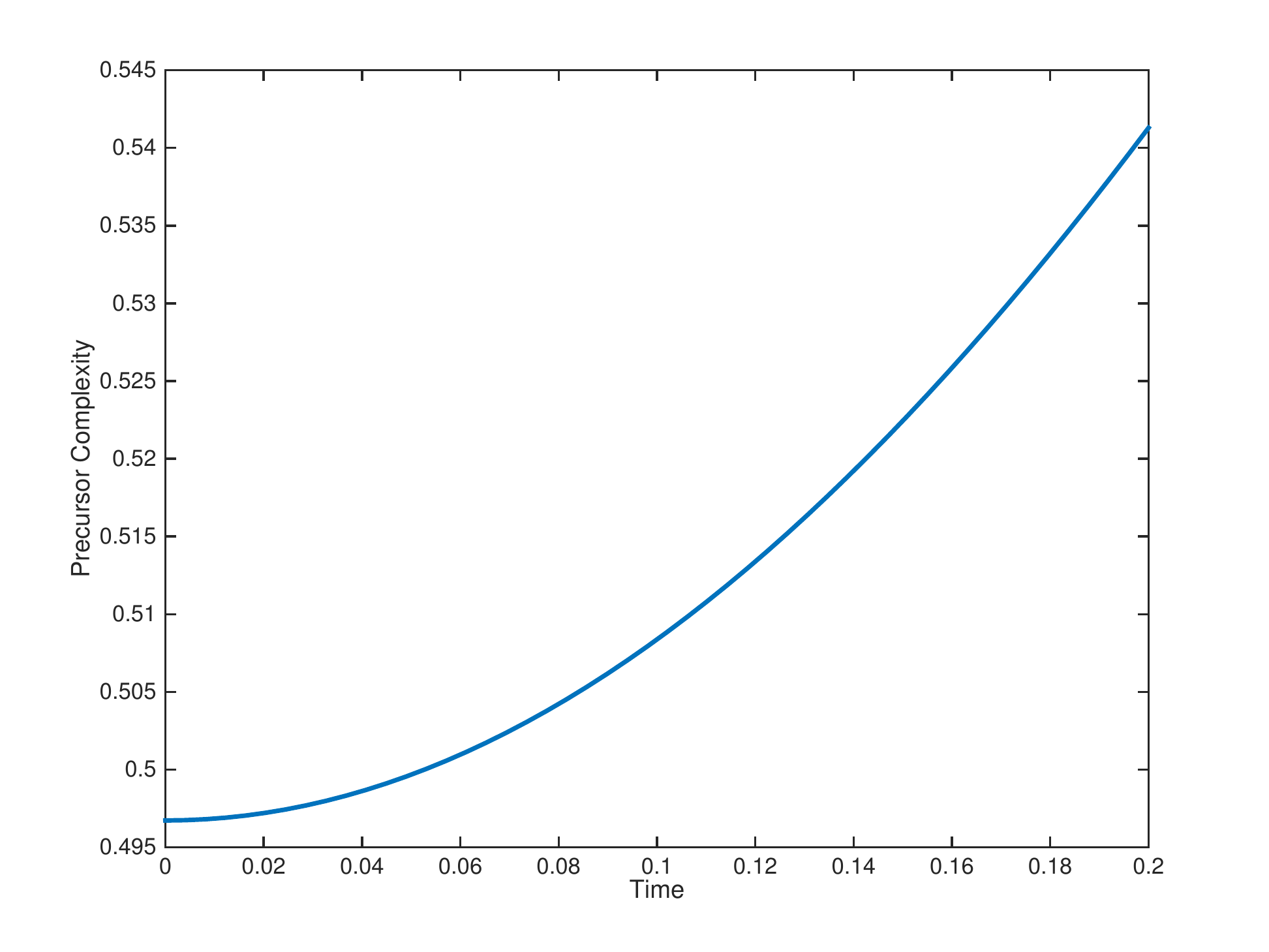}
 \caption{$W_0=iX.$}
\end{subfigure}
\begin{subfigure}{.45\textwidth}

 \includegraphics[width = \textwidth]{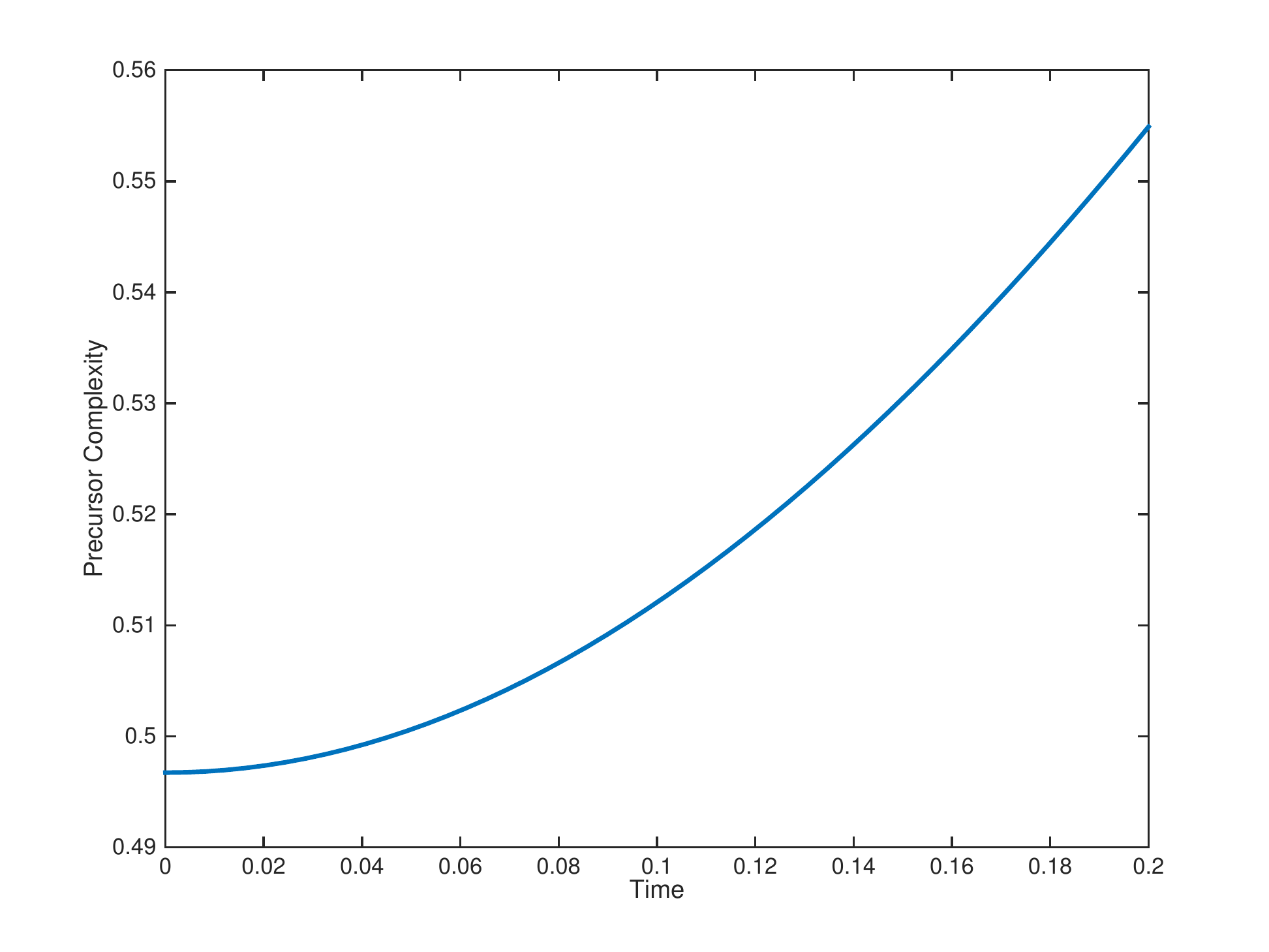}
 \caption{$W_0=iY.$}
\end{subfigure}
  \caption{Complexity of a precursor operator versus time, for the one-qubit system described in the main text, for different choices of the operator $W_0$. The time axis is in units of seconds.}\label{fig:Prec_one_qubit}
\end{figure}

Now we consider a precursor, $W(t)$, which will be a time-evolved ``easy'' operator. In our one-qubit case, we will  consider the operators
$$W(t) = e^{-iHt} iX e^{iHt}$$
and 
$$W(t) = e^{-iHt} iY e^{iHt},$$
where we have chosen the normalization so that $\det W(t) = 1$ and $W(t) \in \text{SU}(2)$.
We numerically solve the geodesic equations, and plot the complexity $\mathcal{C} [ W(t) ]$ as a function of time in Figure~\ref{fig:Prec_one_qubit}, using the same parameters as Figure~\ref{fig:Unit_one_qubit}.

When $t=0$, $W(t) = W_0$ is entirely along an easy direction. However, for nonzero $t$, we will of course have terms proportional to the commutator $[W_0,J_1 X + J_2 Y] \supset Z$, which is a hard direction. In this case, the hardness of $Z$ generates negative curvature in the complexity geometry and is somewhat reminiscent of the switchback effect for large-$N$ systems, as discussed above. That is, the precursor will be a sum of nested operators which are initially suppressed for small times, but become more important for large times. The nested commutators for the large-$N$ case generate terms that are very complex. Hence, the complexity is initially suppressed, and then begins to grow. Indeed, we see that the rate of growth of $\mathcal{C} [ W(t) ]$ is somewhat suppressed for some time, and then it begins to grow linearly, though of course the difference between the two regimes is less stark than we expect in the large-$N$ case. Therefore, it appears that this negative-curvature model of one-qubit complexity is a very simple toy model of information scrambling. 

To see the connection between negatively-curved geometry and the structure of the commutators more concretely, consider the following argument of Brown and Susskind~\cite{SingleQubit}. Say that we have two ``easy'' directions, $\mathcal{O}_1$ and $\mathcal{O}_2$. Then the Baker-Campbell-Hausdorff formula tells us the following: 
$$e^{i \mathcal{O}_1 t} e^{i \mathcal{O}_2 t} = \exp [ i(\mathcal{O}_1+ \mathcal{O}_2)t - \frac{1}{2} [\mathcal{O}_1,\mathcal{O}_2] t^2 + \cdots ].$$
We can think of the LHS as starting from the identity matrix $I$, and then traveling a distance $t$ in the $\mathcal{O}_1$ direction to a point $P$, and then a distance $t$ along the $\mathcal{O}_2$ to a point $Q$. The RHS is taking a path from the identity directly to the point $Q$. If the commutator $[\mathcal{O}_1,\mathcal{O}_2]$ is hard, then the length of the ``hypotenuse'' (the direct path from the identity matrix to the point $Q$) will be longer than what it would be in flat space. This is a generic property of negatively-curved geometries. On the other hand, if the commutator $[\mathcal{O}_1,\mathcal{O}_2]$ is easy, then the length of the ``hypotenuse'' will be shorter than what it would be in flat space. This is a property of positively curved spaces. Indeed, in the present setup, with $\mathcal{I}_{XX} = \mathcal{I}_{YY} = 0.1$, we can calculate, for example, the scalar curvature at the origin (the identity matrix), which is given by 
$\mathcal{R} = 10(8-20 \mathcal{I}_{ZZ})$~\cite{SingleQubit}. Hence, the (curvature of the) complexity geometry will be negative when the $Z$ gate is much harder than the other directions and positive when the $Z$ gate is easier.

\begin{figure}

\centering
\begin{subfigure}{.45\textwidth}

 \includegraphics[width = \textwidth]{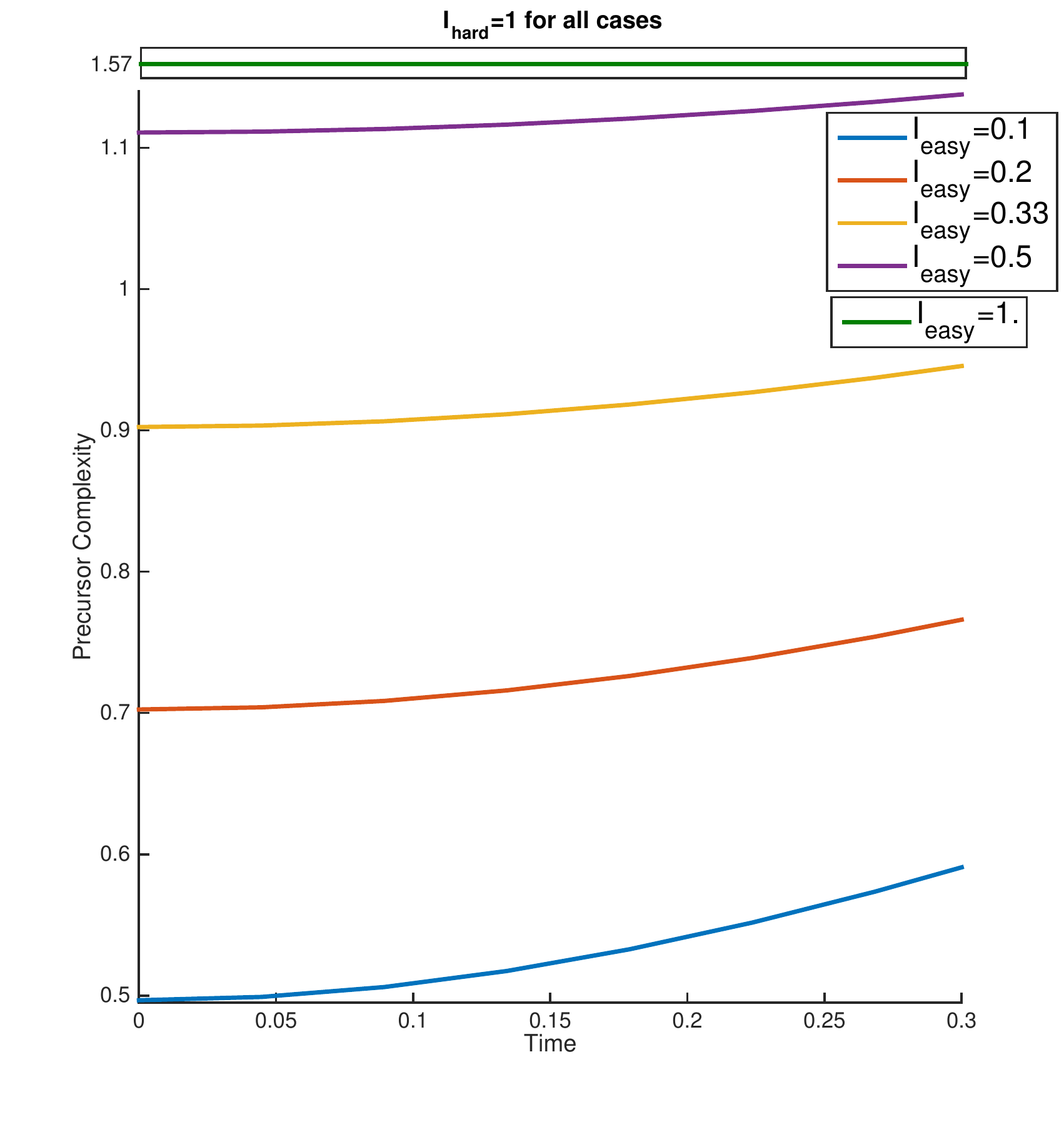}
 \caption{$W_0=iX.$}
\end{subfigure}
\begin{subfigure}{.45\textwidth}

 \includegraphics[width = \textwidth]{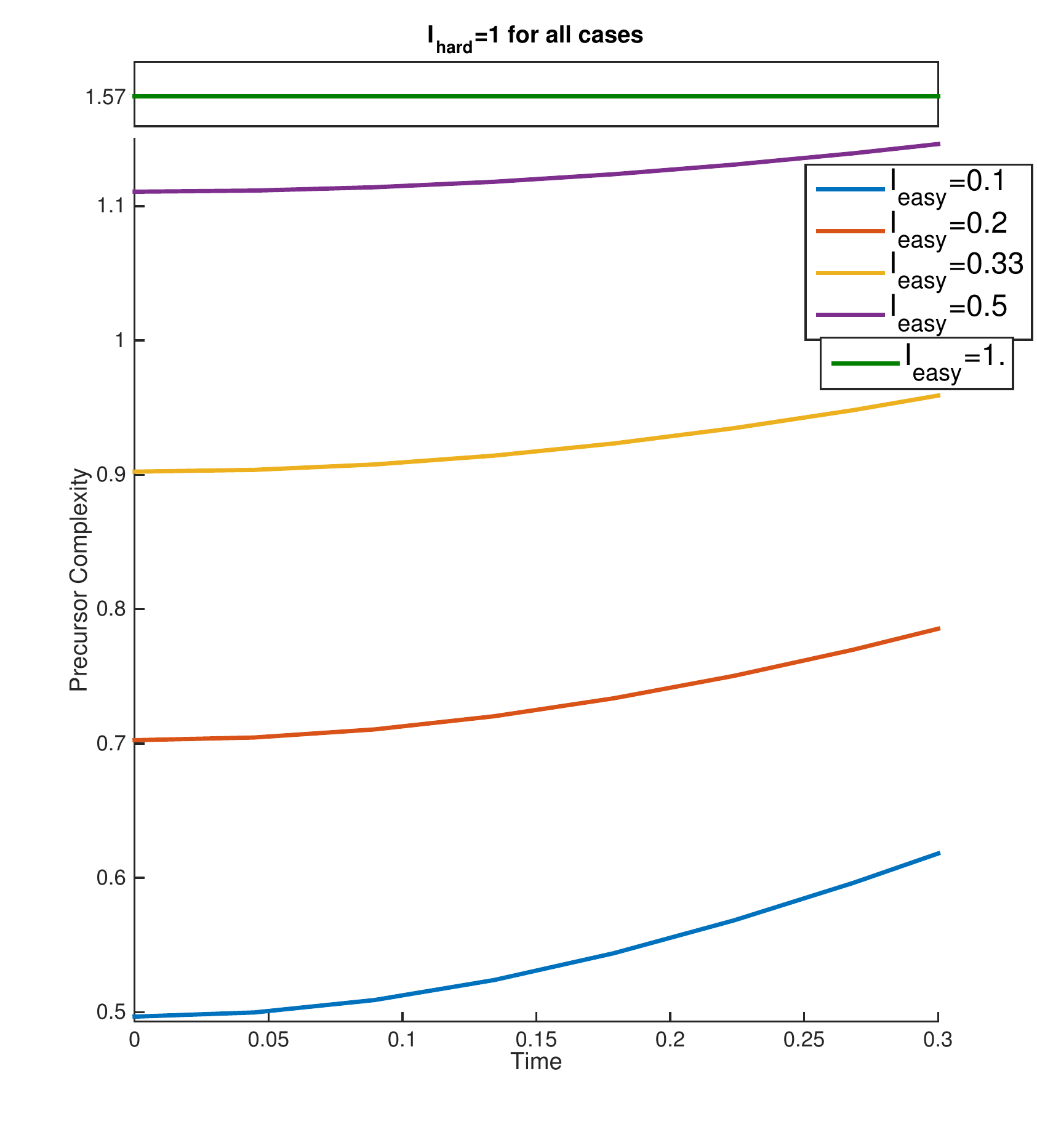}
 \caption{$W_0=iY.$}
\end{subfigure}
  \caption{Complexity of a precursor operator versus time, for the one-qubit system described in the main text, for different choices of the operator $W_0$ for various penalty factors. The very lowest curves in each panel correspond to the plots shown in Figure~\ref{fig:Prec_one_qubit}. The time axis is in units of seconds.  Note that there is a jump in the vertical axis.}\label{fig:Prec_one_qubit_pen_var}
\end{figure}

Finally, we plot the complexity of the same precursors considered above (with the same Hamiltonian) with various penalty factors, in Figure~\ref{fig:Prec_one_qubit_pen_var}. We can see that as we decrease the ``easy'' penalty factor relative to the hard one, the switchback effect becomes more pronounced, as expected. Indeed, for the case where both penalty factors are 1 (i.e., the metric is isotropic), the complexity is a constant as a function of time.  We explain this last case in detail analytically in the next section. Therefore, we see explicitly that this anisotropy in complexity space that generates the negative curvature is essential to the switchback-like-behavior in precursor operators. 

Hence, we see that precursors are a more sensitive probe of the complexity geometry, and in particular, the anisotropy, since their qualitative behavior is affected by the introduction of certain types of anisotropies, as we have shown.  Indeed, the unitary time-evolution operator will be entirely in the easy subspace, and so will not be affected by this type of anisotropy. \footnote{An explicit discussion of the different features of the Nielsen metric probed by precursor and time-evolution unitary complexity is given in appendix~\ref{app:A}.}

\section{Hard Two-Qubit Gates, Easy One-Qubit Gates}

Generally, when one introduces measures of complexity, gates that act on a small numbers of qubits are considered ``easy,'' while those that act on larger numbers of qubits are considered ``hard.'' Thus, when considering a model of circuit complexity for two-qubit systems, one expects that the natural choice is to assign gates that act non-trivially on both qubits a higher penalty factor than those that act on only one qubit. However, we will see in this section that this is not suitable as a model for the switchback effect. In fact, the reverse (i.e., assigning one-qubit gates a much larger penalty factor than their two-qubit counterparts) seems to be a geometry that much better illustrates the switchback effect. This latter model is discussed in the next section.

\begin{figure}
\centering
 \includegraphics[width = 0.75\textwidth]{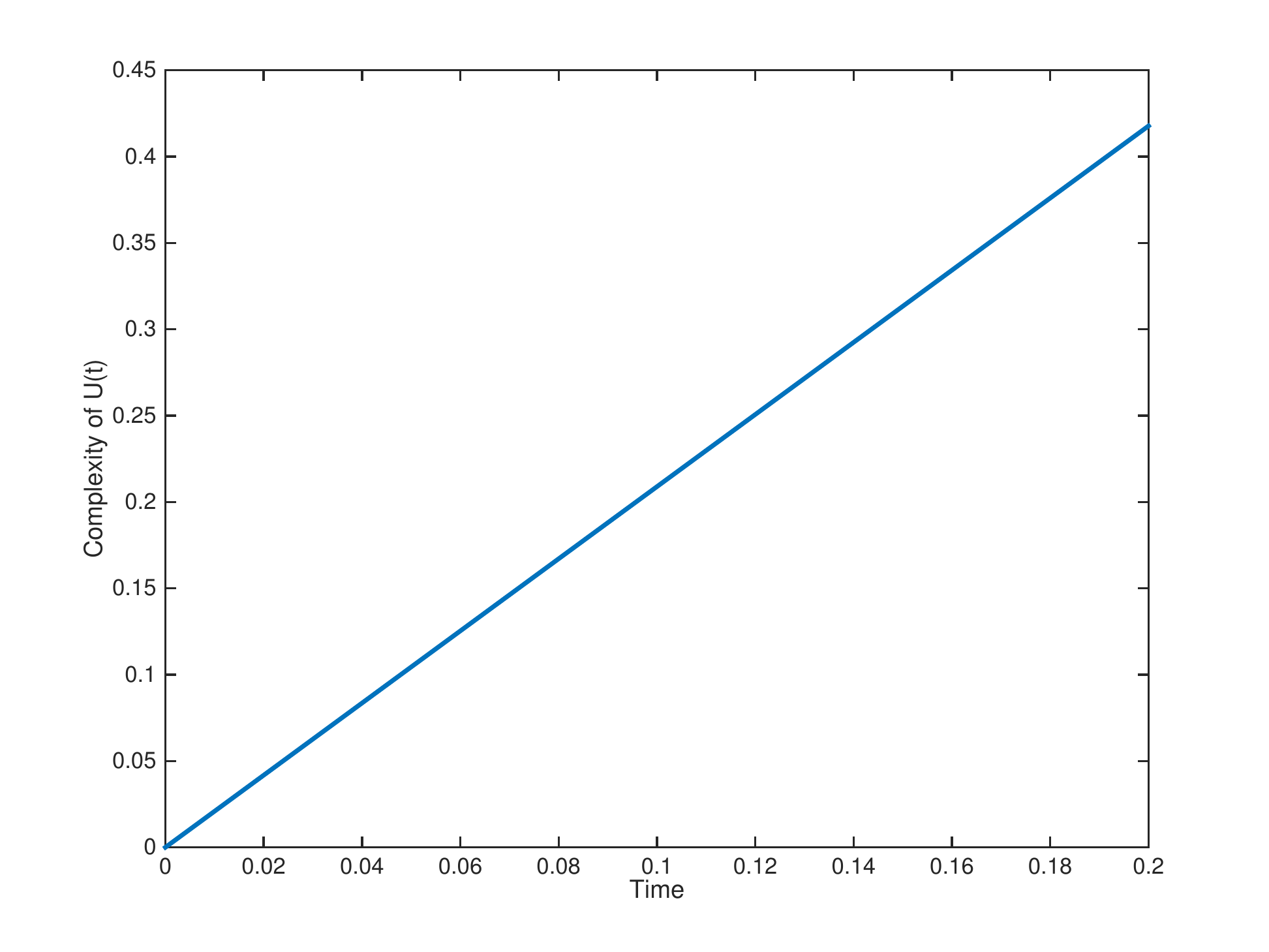}

  \caption{Complexity of a unitary time-evolution operator versus time with one-qubit gates being easier than two-qubit gates. The complexity of this operator grows linearly with time. The time axis is in units of seconds.}\label{fig:Unit_one}
\end{figure}

Consider a precursor operator in this model with a large cost factor for the two-qubit gates. In this case, our ``local'' Hamiltonian is one built out of easy, i.e., one-qubit operators. In general, it will take the form
$$H = \sum_i \left ( J_{1i}~\sigma_i \otimes I  + J_{2i}~ I \otimes \sigma_i \right ).$$ 
We numerically solve the geodesic equations to calculate the complexity of the unitary time evolution operator $U(t)$ corresponding to this type of Hamiltonian (using the geometry described previously). We plot the results in Figure \ref{fig:Unit_one}. Explicitly, the cost factor will be taken to be $0.1$ for one-qubit gates, and $1.0$ for two-qubit gates. We use the following coefficients in the Hamiltonian:
$$J_{1X} =   0.9390, J_{1Y} =0.8759,J_{1Z} =0.5502,J_{2X} =0.6225 , J_{2Y} =0.5870,J_{2Z} =0.2077,$$
where each $J_{ij}$ is given in units of $s^{-1}$. We see that it is linear in time for early times, as expected. 

Recall that a precursor takes the form 
$$W(t) = e^{-i H t} W_0 e^{iH t},$$
where $W_0$ is an ``easy'' operator, a one-qubit operator in this case. As the Hamiltonian is a sum of commuting single-body terms we can clearly see that $W(t)$ will also be a single-body operator as the evolution does not generate any operators on the second qubit. Say for concreteness that $W_0 = i X \otimes I$ (the overall factor is chosen so that $W(t)$ has determinant 1). Then our precursor $W(t)$ becomes 
\begin{equation}
\begin{aligned}
W(t) & = \exp( - i \sum_i \left ( J_{1i}\sigma_i \otimes I  + J_{2i} I \otimes \sigma_i \right )t) iX \otimes I \exp( i \sum_i \left ( J_{1i}\sigma_i \otimes I  + J_{2i} I \otimes \sigma_i \right )t)  \\ & = \exp( - i \sum_i \left ( J_{1i}\sigma_i  \right )t) iX \exp( i \sum_i \left ( J_{1i}\sigma_i   \right )t) \otimes I.
\end{aligned} 
\end{equation} 
As we are now only considering single-qubit operators, the matrix exponentials can be easily computed giving schematic form of (for example) 
\begin{equation}
\begin{aligned}
W(t) & = \exp( i \sum_i \alpha_i \sigma_i ) iX \exp( -i \sum_i \alpha_i \sigma_i ) \\
\\ & = \left ( \cos \left (\alpha \right ) I+ i \sum_i \hat \alpha_i \sigma_i \sin \alpha \right ) iX \left ( \cos \left (\alpha \right )I - i \sum_i \hat \alpha_i \sigma_i \sin \alpha \right ),
\end{aligned}
\end{equation}
where $\alpha \equiv \sqrt{ \alpha_1^2 + \alpha_2^2 + \alpha_3^2},$ and $\hat \alpha_i \equiv \alpha_i / \alpha$, are $t$-dependent parameters.  Using the identity 
$$\sigma_i \sigma_j = I \delta_{ij} + i \epsilon_{i j k} \sigma_k,$$
we can show that $W(t)$ has no term proportional to the identity, since, by cyclicity of trace, $\text{Tr} W(t) = \text{Tr} W(0) = 0.$ 
\begin{figure}
\centering
 \includegraphics[width = 0.75\textwidth]{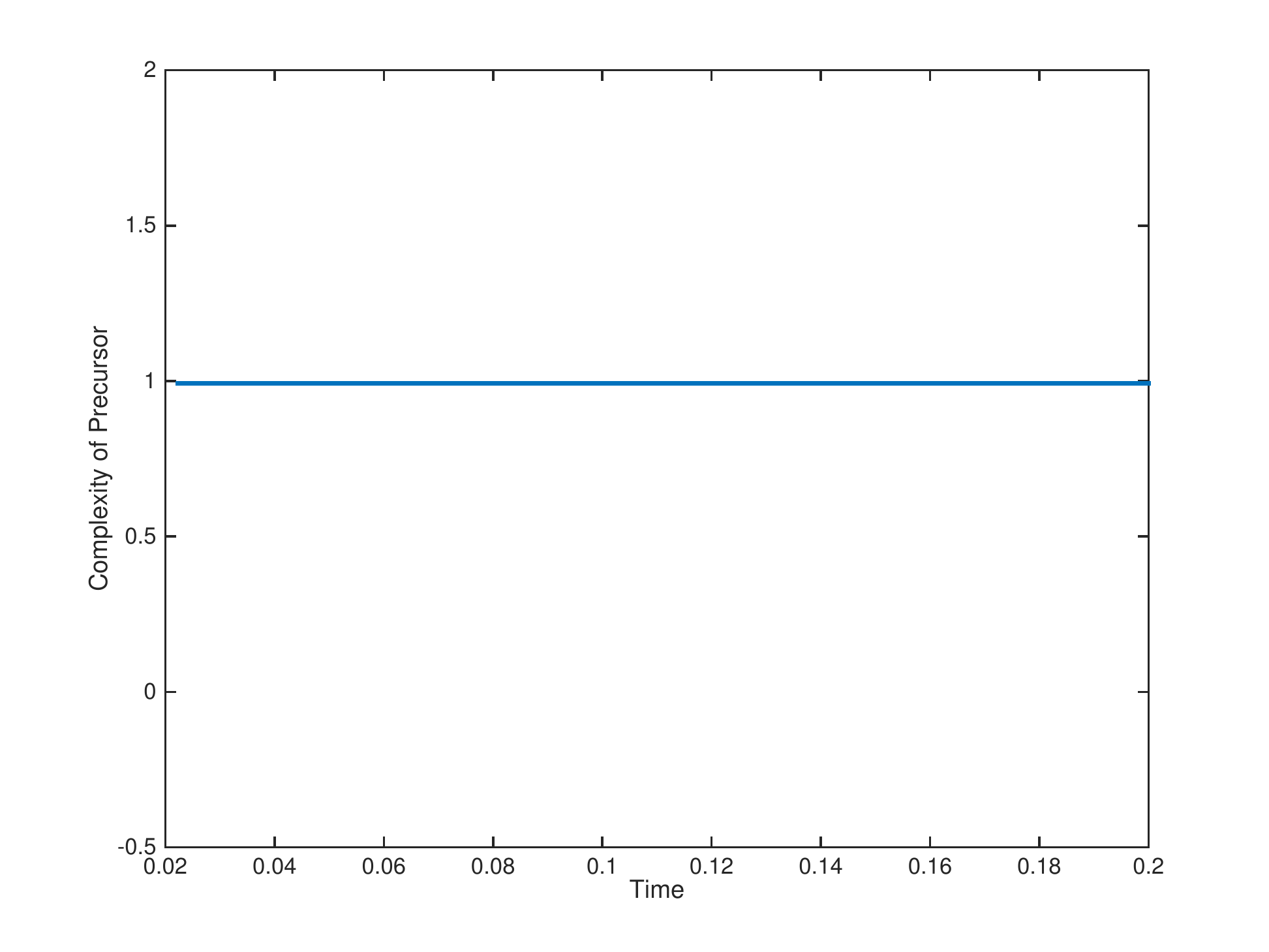}

  \caption{Complexity of a precursor operator versus time, with one-qubit gates being easier than two-qubit gates. In this case, $W_0 = i X \otimes I.$ The complexity of the precursor is constant in time, as explained in the main text. The time axis is in units of seconds.}\label{fig:Prec_one}
\end{figure}


 Therefore, for all time $t$, the precursor $W(t)$ will have no term proportional to the identity, since all the Pauli matrices are traceless. We can write $W(t)$ in the form 
$$W(t) =\exp( i \sum_i \beta_i(t) \sigma_i).$$
We can expand this as
$$W(t) =   \exp( i \sum_i \beta_i(t) \sigma_i)=  \left ( \cos \left (\beta(t) \right ) I+ i \sum_i \hat \beta_i(t) \sigma_i \sin \beta (t) \right ),$$
where $\beta(t) \equiv \sqrt{ \beta_1(t)^2 + \beta_2(t)^2 + \beta_3(t)^2},$ and $\hat \beta_i(t) \equiv \beta_i(t) / \beta(t)$.\footnote{See appendix~\ref{app:A} for an explicit calculation of $\beta_i(t)$.}

Because $W(t)$ has no term proportional to the identity, we must have $\cos \beta = 0$ so that $\beta = \frac{\pi}{2}$ for all time. In this geometry, all one-qubit directions have the same cost, so we would expect that the complexity of the precursor will be constant throughout time. Once again, we solve the geodesic equations, and plot the complexity of $W(t)$ versus time in Figure \ref{fig:Prec_one}. We see that $\mathcal{C} [ W(t) ]$ is constant with respect to time. (Note that this is exactly the same reason why the complexity of a precursor was constant in time for the one-qubit case with an isotropic metric discussed at the end of the last section.) Although we chose $W_0=i X \otimes I$ for concreteness, the complexity of $W(t)$ will be constant in time for any choice of one-qubit operator $W_0$ with $\det W_0 = 1$, using the exact same logic as above. However, different choices of $W_0$ will result in different values for the complexity.  Hence, this geometry is poorly-suited to model the switchback effect.  Indeed, as discussed above, since the commutator of two ``easy'' directions is ``easy,'' the complexity geometry is not negatively-curved and we should not expect the precursor to display switchback-effect-like behavior.

We now turn to the other case (where the two-qubit gates have a lower cost factor than the one-qubit gates), which seems to be a better model of the switchback effect. 


\section{Easy Two-Qubit Gates, Hard One-Qubit Gates}

\begin{figure}
\centering
 \includegraphics[width = 0.75\textwidth]{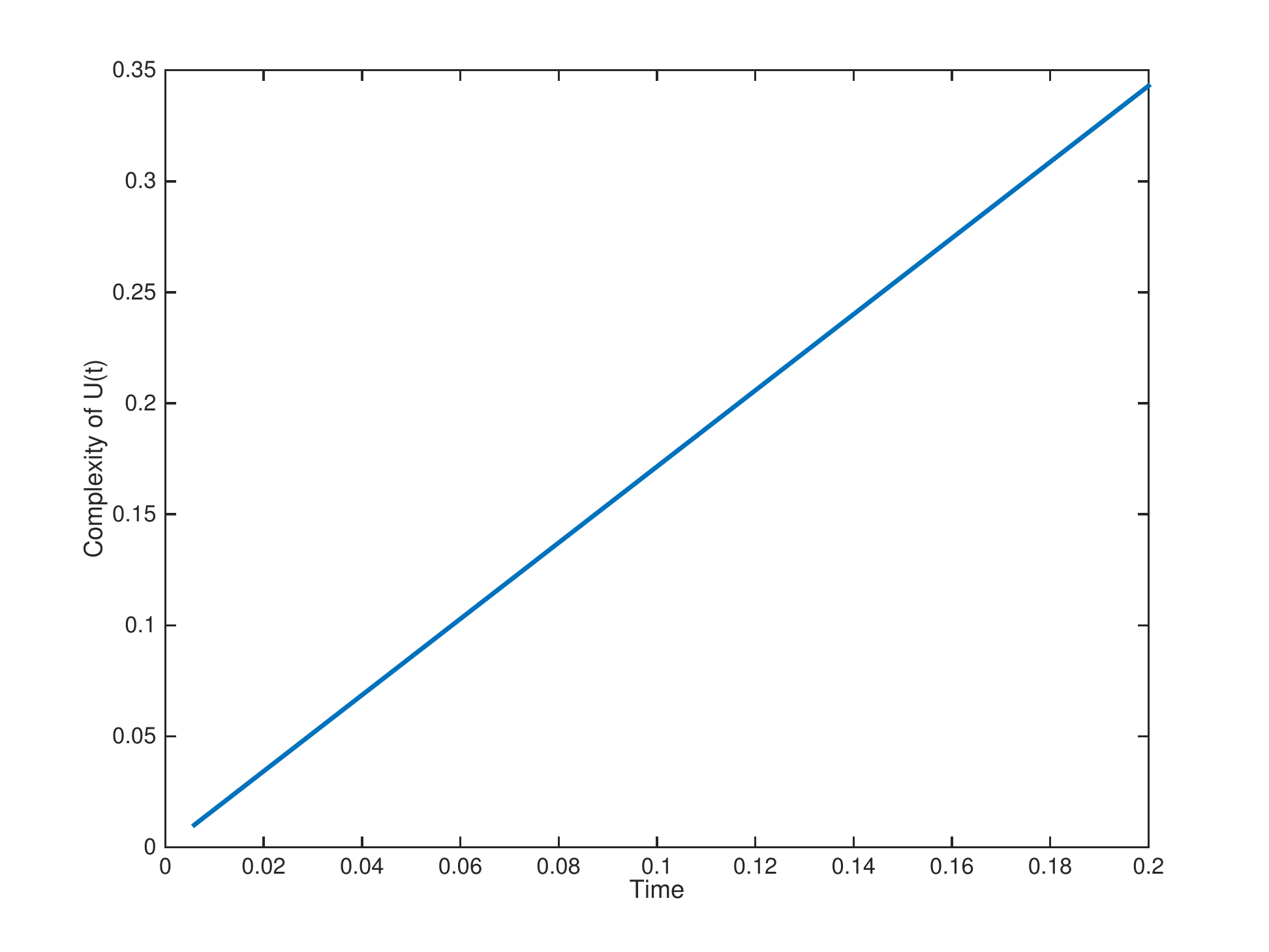}

  \caption{Complexity of a unitary time-evolution operator versus time with one-qubit gates being harder than two-qubit gates. The complexity of this operator grows linearly with time. The time axis is in units of seconds.}\label{fig:Unit_inv}
\end{figure}

In this section, we consider a geometry in which the two-qubit gates are ``easy'' while the one-qubit gates are ``hard.'' While this may seem counter-intuitive from the perspective of large-$N$ systems (where gates that act on a small number of qubits are easy), we shall see that the geometry that results from this choice seems to illustrate the switchback effect. In this case, the Hamiltonian will be a sum of ``easy'' terms, 
$$H = \sum_{ij} J_{ij}~ \sigma_i \otimes \sigma_j.$$

The unitary time-evolution operator is then $U(t) = \exp (-iHt)$. We solve the geodesic equation numerically, and use this to calculate the complexity $\mathcal{C} [ U(t)]$. We plot the results in Figure \ref{fig:Unit_inv}, for Hamiltonian 1 in Table~\ref{HamiltonianTableTwoQubits}. The Hamiltonians in this table were found by randomly generating the nine two-qubit operator coefficients. As expected, the complexity grows linearly in time. 

\begin{figure}
\begin{subfigure}{0.45\textwidth}
\centering
 \includegraphics[width = \textwidth]{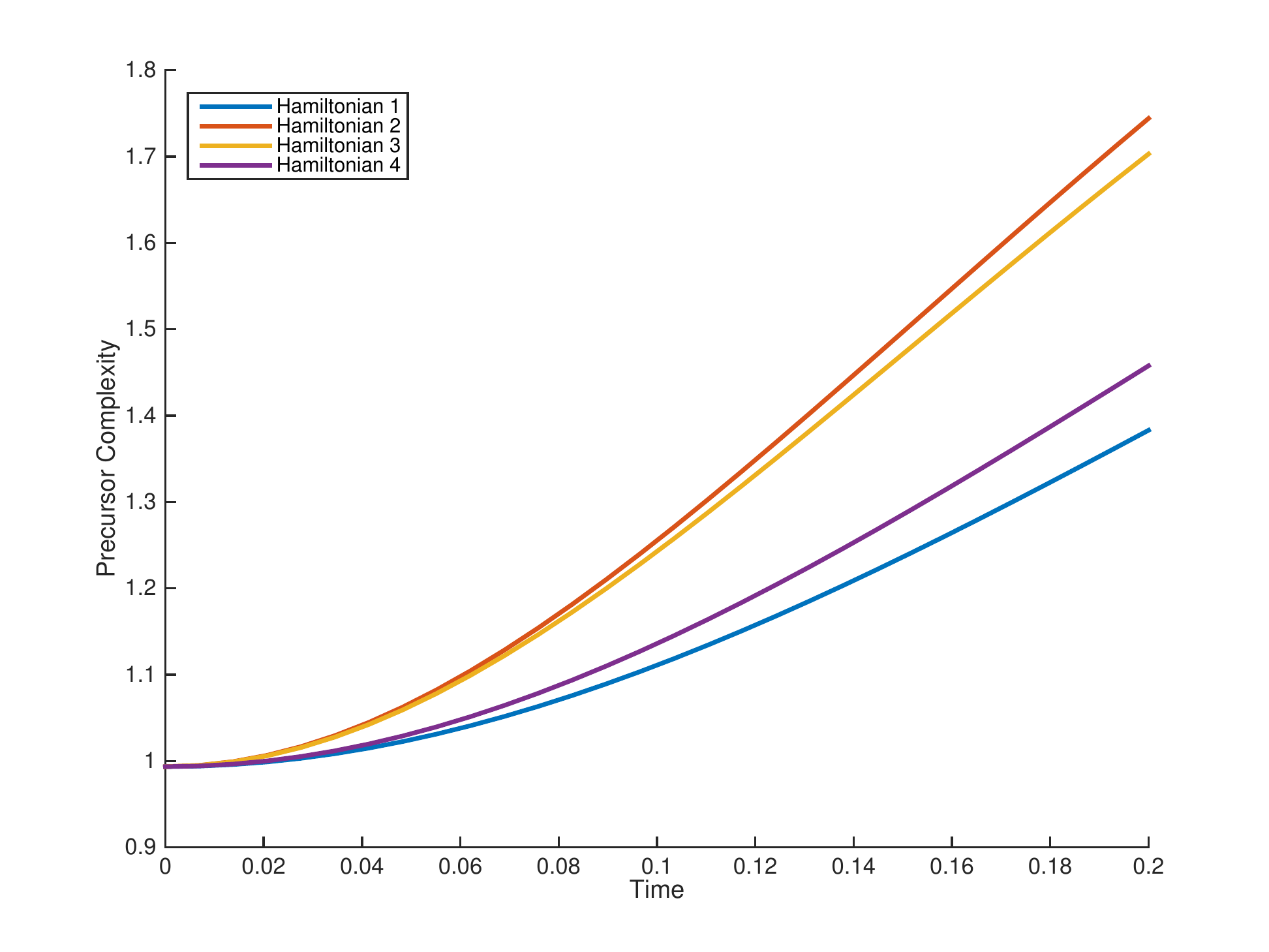}
  \caption{$W_0 = - X \otimes X$}
 \end{subfigure}
 \begin{subfigure}{0.45\textwidth}
\centering
 \includegraphics[width = \textwidth]{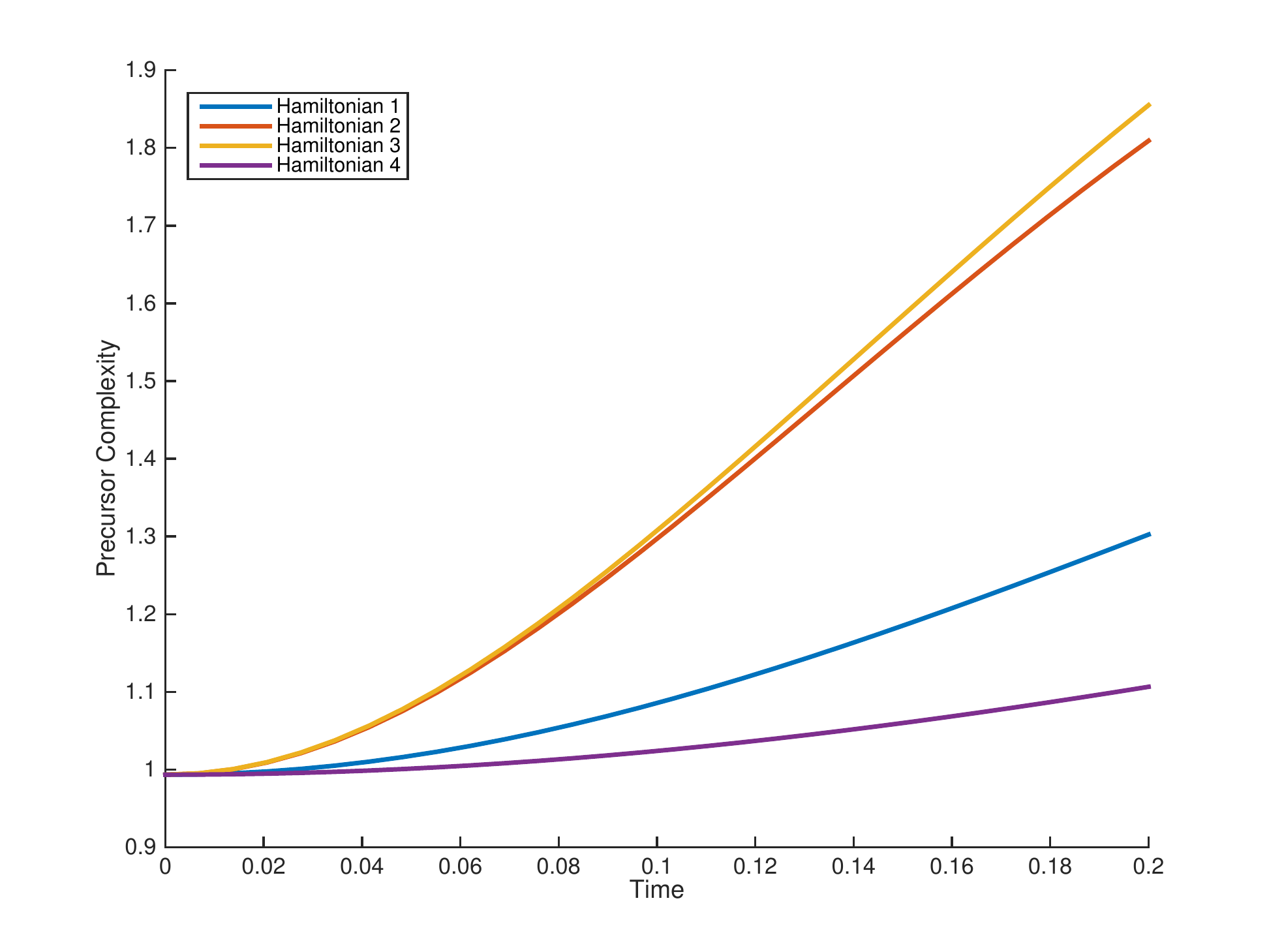}
 \caption{$W_0 = - X \otimes Y$}
 \end{subfigure}
 
  \begin{subfigure}{0.45\textwidth}
\centering
 \includegraphics[width = \textwidth]{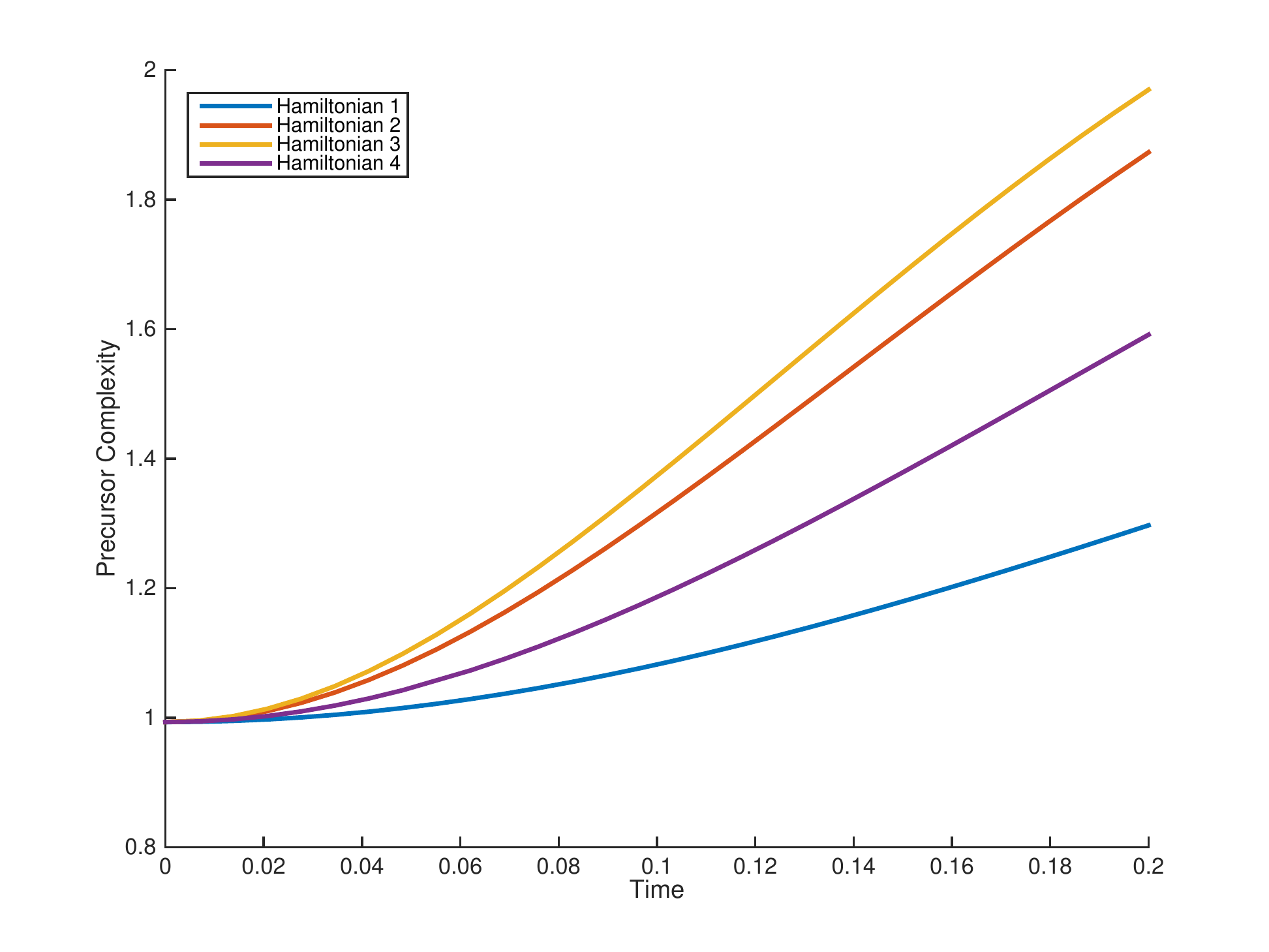}
 \caption{$W_0 = - X \otimes Z$}
 \end{subfigure}
  \begin{subfigure}{0.45\textwidth}
\centering
 \includegraphics[width = \textwidth]{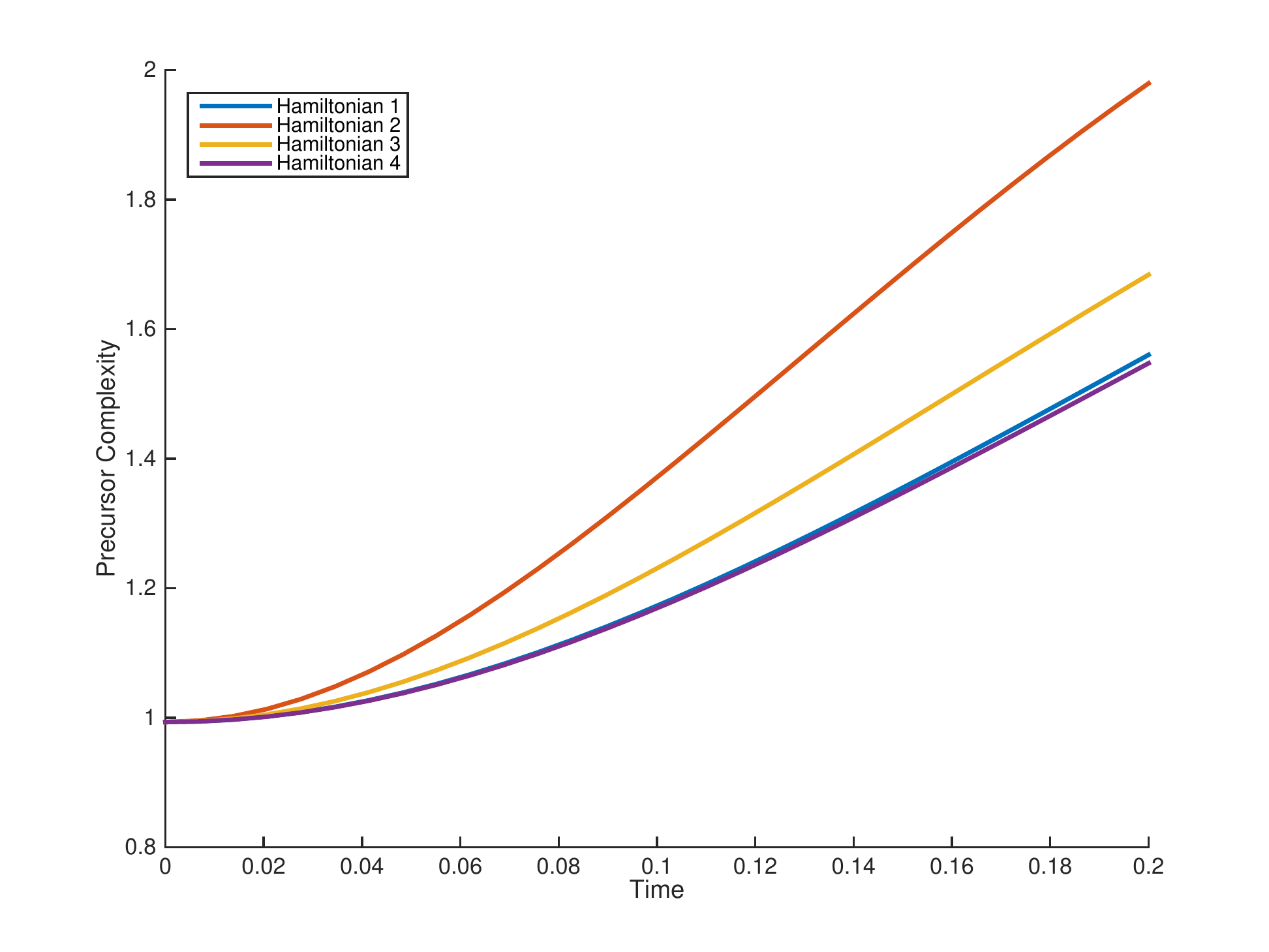}
 \caption{$W_0 = -Y \otimes Z$}
 \end{subfigure}

  \begin{subfigure}{0.45\textwidth}
\centering
 \includegraphics[width = \textwidth]{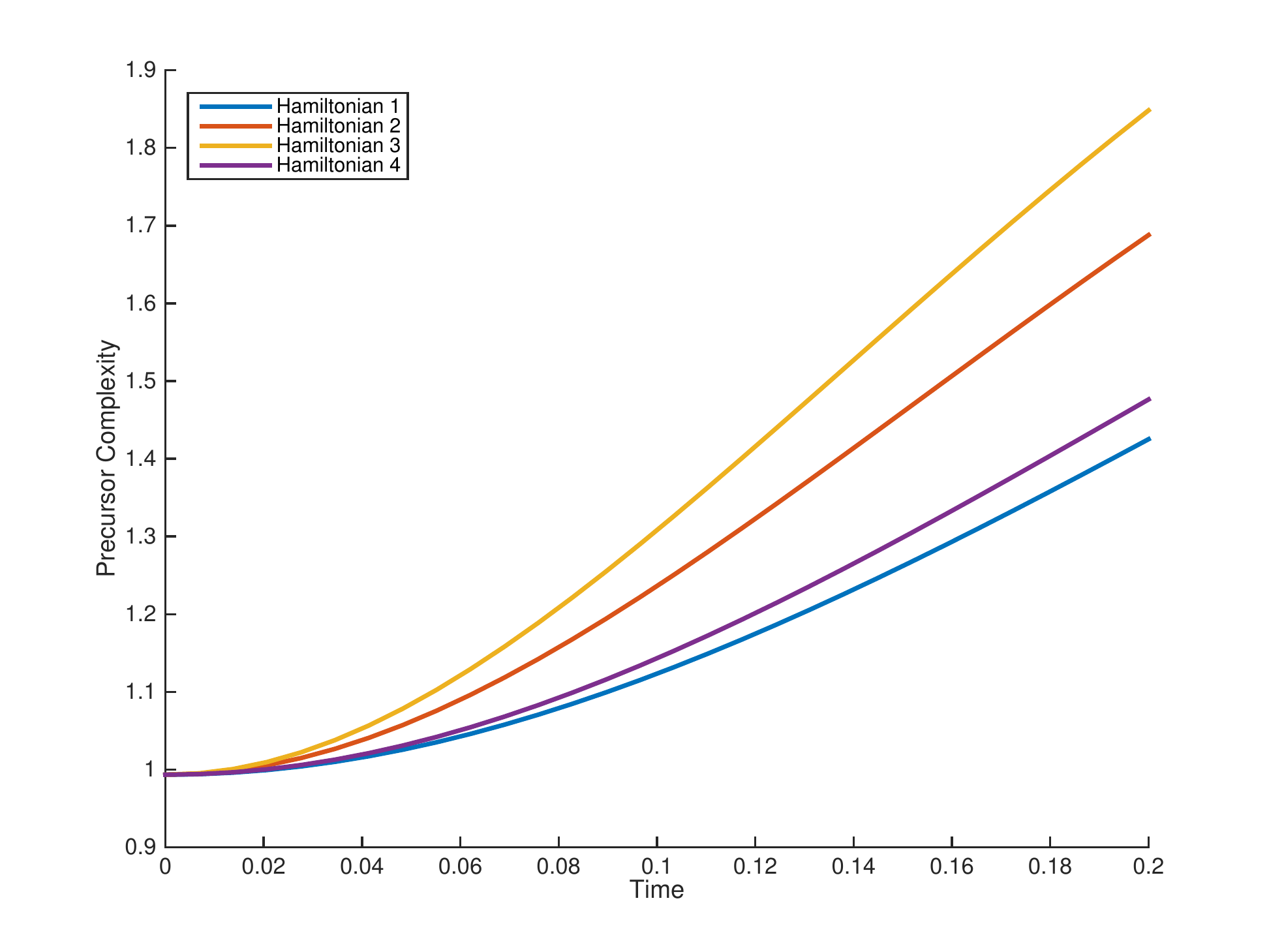}
 \caption{$W_0 = - Y \otimes Y$}
 \end{subfigure}
  \begin{subfigure}{0.45\textwidth}
\centering
 \includegraphics[width = \textwidth]{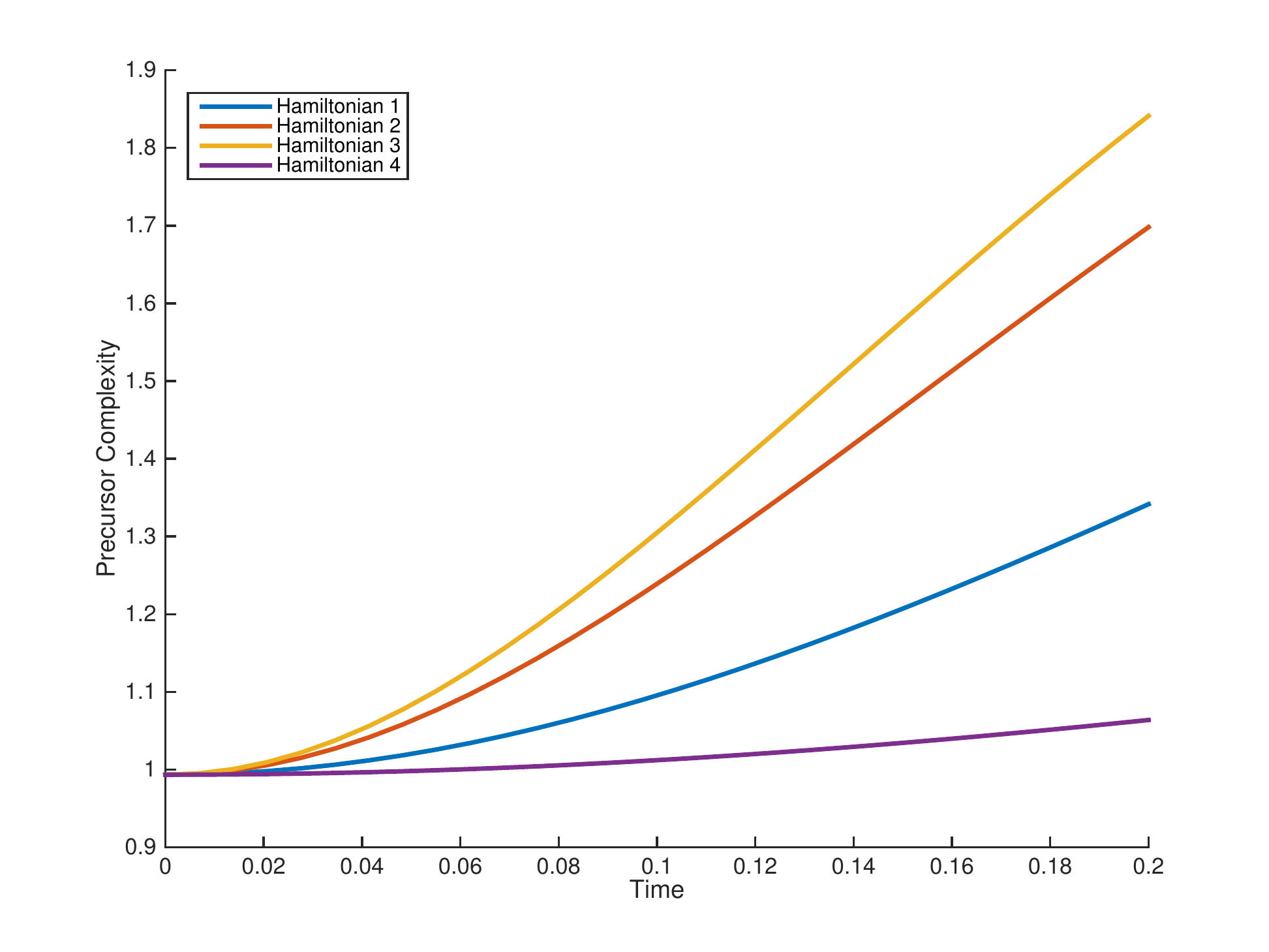}
 \caption{$W_0 = - Z \otimes Z$}
 \end{subfigure}
  \caption{Complexity of a precursor operator versus time, with one-qubit gates being harder than two-qubit gates for various choices of the operator $W_0,$ as well as for various choices of the Hamiltonian. The complexity of the precursor grows slowly at first, then begins to grow linearly. This is a result of a switchback effect similar to the one seen in systems with a large number of degrees of freedom. The time axis is in units of seconds. Note that the overall factor is chosen so that $\det W(t) = 1$.}\label{fig:Prec_inv}
\end{figure}

Recall that a precursor in this geometry will have the form 
$$W(t) = e^{-iHt} W_0 e^{iHt},$$
however now $W_0$, being ``easy,'' will of course act non-trivially on both qubits. At $t=0$, this operator will only have non-zero coordinates in an ``easy'' direction. However, as we increase $t$, we will have terms in $W(t)$ that result from the commutator of terms in the Hamiltonian and $W_0$. In general, these will lead to terms that act non-trivially on only one qubit. For example, if $W_0 =  -X \otimes X$, $ [W_0, X \otimes Y]$ will be proportional to $I \otimes Z,$ which is a ``hard'' gate. Note that this is a somewhat similar mechanism to what happens to the complexity of precursors in large-$N$ systems. As discussed above, this structure of the commutator (specifically that the commutator of two ``easy'' operators yields a ``hard'' operator) is a reflection of the negative curvature of unitary space that is a key component of switchback-like time evolution of precursor complexity. This should be contrasted with the situation above, where the ``easy'' gates were taken to be the one-qubit operators. In this case, the commutator of two ``easy'' gates is another ``easy'' gate. Therefore, even though \textit{physically} (in, for example, a laboratory setting) this setup (in which two-qubit gates are taken to be easier than one-qubit) gates is not a good model of complexity, it is a good mathematical toy model of the switchback effect. In particular, once again, we see that the negative curvature is necessary to generate switchback-like-behavior of precursor operator complexity. 

\begin{table}
\begin{tabular}{ |p{1.45cm}||p{1.cm}|p{1.cm}|p{1.cm}|p{1.cm}|p{1.cm}|p{1.cm}|p{1.cm}|p{1.cm}|p{1.cm}|}
 
 \hline
   & $J_{11}$  & $J_{12}$  & $J_{13}$ & $J_{21}$  & $J_{22}$  & $J_{23}$ & $J_{31}$  & $J_{32}$  & $J_{33}$ \\
 \hline
$H_1$   & 0.1888    & 0.0012&   0.3164&   0.6996&   0.6253&   0.5431 & 0.4390 &0.2874 & 0.5017\\
\hline
$H_2$ & 0.9649    &0.1576    &0.9706    &0.9572   & 0.4854    &0.8003    &0.1419    &0.4218  &  0.9157 \\
\hline
$H_3$ & 0.7922   & 0.9595    &0.6557 &   0.0357 &   0.8491 &   0.9340  &  0.6787   & 0.7577 &   0.7431 \\
\hline
$H_4$ & 0.3922    &0.6555  &  0.1712   & 0.7060   & 0.0318   & 0.2769  &  0.0462  &  0.0971  &  0.8235 \\
 \hline

\end{tabular} 
\caption{Hamiltonians considered in the main text. Each row in the table corresponds to $H= \sum_{ij} J_{ij}~ \sigma_i \otimes \sigma_j.$ Each $J_{ij}$ is in units of $s^{-1}$.}
\label{HamiltonianTableTwoQubits}
\end{table}

When we calculate the complexity of $W(t)$ and plot the results as a function of time (see Figure \ref{fig:Prec_inv}), we see that for a brief initial period, the complexity grows slowly, and then it begins to grow much more rapidly, exactly as expected from the previous discussion. We do this calculation for the Hamiltonians given in Table~\ref{HamiltonianTableTwoQubits}. This behavior of complexity of precursors is very similar in form to the behavior of precursors in the usual systems with a much larger number of qubits~\cite{QCNC,Switchback}, suggesting that, qualitatively, the switchback effect is related to the generation of ``hard'' operators via commutators of ``easy'' operators and the role of large $N$ is simply to increase the number of ``hard'' operators relative to ``easy'' operators, sharpening the effect.

\section{Conclusions and Future Directions}

In this paper, we have analyzed complexity of operators in one- and two-qubit systems, using Nielsen complexity geometry. This approach allows us to give a precise definition of operator complexity, and largely confirm our qualitative intuition for how this should behave. We have found that the unitary time-evolution operators grow linearly with time, as expected. It was also found that precursor operators grow more slowly at first, at least when we chose our ``hard'' and ``easy'' gates such that the resulting complexity geometry was negatively curved. For one-qubit systems, this occurred when we considered one of the three Pauli operators to be ``hard,'' and the other two to be ``easy.'' For two-qubit systems, to obtain negative curvature, we chose two-qubit gates to be ``easy,'' while considering their one-qubit counterparts to be ``hard.''  While this is somewhat counter-intuitive from, e.g., a laboratory point of view, this choice was necessary to generate negative curvature that is required to model many of the features of computational complexity. 

There are several interesting ways in which our analysis could be extended, which we leave to future work. First of all, it would be worthwhile to try to extend these types of computations to larger numbers of qubits. In particular, it would be nice to consider a system of $N$ qubits, where the one- and two-qubit gates are much less difficult than gates acting on larger numbers of qubits, reflecting laboratory conditions.  In this setting it is natural to consider two-qubit operators ``easy'' and commutators of ``easy'' operators will generate larger ``hard'' operators. As evidenced by our two-qubit calculations, the production of ``hard'' operators (one-qubit in that case) by commutators of ``easy'' operators seems to be an important ingredient needed to observe (qualitatively) the switchback effect. We therefore expect that the behavior of precursor complexity will look more and more like the behavior expected for the large-$N$ limit--very small complexity until the scrambling time, and then linear growth. 

Furthermore, it would be interesting to analyze specific qubit models, especially those that are of interest from the gravity point of view, such as the SYK model. For, e.g., the SYK model, the scrambling time is expected to scale logarithmically with the number of degrees of freedom. Presumably, then, these calculations would show that the time at which linear growth starts also scales logarithmically with the number of degrees of freedom. 

However, doing these calculations for a larger number of degrees of freedom means analyzing a much more complicated geometric space, the dimension of which scales exponentially in the number of qubits, which poses a considerable computational challenge.  

We expect that future work analyzing these types of problems will continue to illuminate the connections between information, holography, gravity, and spacetime, and in particular illustrate how bulk data is encoded in information-theoretic quantities in its holographic boundary dual.

\acknowledgments

It is a pleasure to thank Vijay Balasubramanian, Adam Brown, Matt DeCross, Hong Liu, Hugo Marrochio, and Shreya Vardhan for valuable comments on the manuscript. We are also grateful to Raphael Bousso, Yasunori Nomura, Pratik Rath, and Vincent Su for fruitful conversations. The work of SL is supported by the U.S. Department of Energy, Office of Science, Office of High Energy Physics of U.S. Department of Energy under grant Contract Number  DE-SC0012567. SL acknowledges the support of the Natural Sciences and Engineering Research Council of Canada (NSERC).

\appendix
\section{Parameterization of One-Qubit Operators and Recurrence Times }
\label{app:A}

In this Appendix, we give an analytic discussion of the parameterization of one-qubit operators and of recurrence times for time-evolution operators. Much of our discussion here overlaps with that in \cite{SYKComp}. 
A general single-qubit ``easy'' Hamiltonian in our setup may be described by a real three-dimensional vector $J_i$ with\footnote{In principle one should add a multiple of identity to $H$ to ensure it is positive definite; however, this will not be important for our discussion.} 
$$ H = \sum_{i=1}^2 J_i~ \sigma^i.$$
Accordingly, we may introduce the magnitude $J = \sqrt{\sum_{i=1}^2 J_i^2}.$ 
The time-evolution unitary is then
$$ U(t) = \cos(Jt) - i\sin(Jt)~ \hat{J}_i \sigma^i$$ 
with $\hat{J}_i \equiv J_i/J$. This makes it clear that the recurrence time-scale of the system is $t_{rec} = 2\pi/J$. It is clear that the coordinates describing the time-evolution unitary are $x^I(t) = - J_I t$ for the easy directions, and $x^3=0$ for the hard direction, since we have $U(t) = \exp(-i J_i \sigma^i t)$. This is certainly true for small times $t << t_{rec}$; however, at large times, we can see that our parameterization of $SU(2)$ `breaks down'.\footnote{That this should happen is clear since, topologically, $SU(2) \sim S^3$ so it should be described by three angles, not an arbitrary three-dimensional vector. In geometric language, the coordinate chart, $x^I$, we have been using is really only valid for a subset of the $x^I$, not for $x^I \in \mathbb{R}^3.$} Thus we see that the unitary described by $x^I(t)$ is equivalent to that described by $\tilde{x}^I_n(t) = - J_I t_n$ with $t_n \equiv t + 2\pi n/J$, for $n \in \mathbb{Z}$, and $I \in \{1,2\}$, $\tilde{x}^3_n = 0.$ This is an example of a `topological obstruction' discussed earlier.\\

It was shown in~\cite{SYKComp} that in the submanifold generated by easy directions, there are linear geodesics $\gamma_n$ between the identity and any point in the ``easy'' submanifold. These geodesics are described by coordinate functions of the parameter $s \in [0,1]$:
$$ x^I_n(s;t) =  - J_I \cdot t_n \cdot s,$$
for $I \in \{1,2\}$, and $x^3_n(s;t)=0$. It is clear from the metric in the space of unitaries, Eq~\ref{eq:metric_def}, that the metric at the identity will be proportional to the cost factor, since our generators $T_I$ are orthogonal. The above solution solves the geodesic equation, so $s$ is an affine parameter, and therefore, the metric projected along the geodesic (expressed in terms of $s$) is a constant along the geodesic, since $x^I \propto s$. The geodesic passes through the identity, so the metric on the one-dimensional space (the geodesic) is the metric projected onto the tangent vector of our geodesic, which we call $h(s)$. This is 
$$h(s) = g_{IJ} \frac{d x^I_n}{ds}\frac{d x^J_n}{ds} = 4 \sum_{I=1}^2 J_I^2 \mathcal{I}_{II} \cdot (t_n)^2.$$

The length of these geodesics in the Nielsen metric are therefore given by 
$$ l(\gamma_n) = 2 \sqrt{\sum_{I=1}^2 J_I^2 \mathcal{I}_{II}} \cdot |t_n|,$$
and minimizing over this family of geodesics gives the complexity 
$$ \mathcal{C} [U(t)] = 2 \sqrt{\sum_{I=1}^2 J_I^2 \mathcal{I}_{II}} \cdot \min_{n \in \mathbb{Z}} |t + 2\pi n/J|.$$ It is clear from this formula that the complexity of the time-evolution unitary will have initial linear growth until $t = \pi/J$ at which point it will decay linearly to zero and then the same curve will repeat with period $2\pi/J$. The geodesics with non-minimal $n$ describe paths that `wind' around the group many times before finally ending at the desired unitary.

The formula above for $\mathcal{C} [U(t)]$ shows us that the unitary complexity always grows linearly with time, and the only feature that is sensitive to the precise details of the complexity geometry is its slope. On the other hand, the qualitative behavior of the precursor complexity is sensitive to the precise details of the complexity metric. 

We now briefly discuss precursor operators. We wish to discuss Pauli operators, however, these have determinant $-1$ and thus are not in $SU(2)$ so we choose to ``re-phase by $i$'' so that the operator $\tilde{\sigma}^i = i\sigma^i$ is described by coordinates $x^j = \frac{\pi}{2} \delta^{ji}$. Conjugation of $\tilde{\sigma}^i$ by the time evolution unitary yields an operator that is rotated by angle $2Jt$ around the $\hat{J}$ axis. Clearly this operator $\tilde{\sigma}^i(t)$ is still described by coordinate vector $x^j(t)$ of magnitude $\pi/2$. The complexity of the precursor is given by the length of the geodesic from the identity to the vector $x^j(t)$ in the Nielsen metric.

One finds that, for an initial operator $W_0 = F_i \tilde{\sigma}^i$ (with $|F|^2 = 1$ for unitary) we find the coordinates of the precursor $W(t)$ to be
$$ \beta^k(t) = \frac{\pi}{2}\left(\cos(2Jt) F_k + (1-\cos(2Jt))(F_l \hat{J}_l) \hat{J}_k + \sin(2Jt) (\epsilon_{ijk} \hat{J}_i F_j)\right).$$

As we mentioned in the main text, the complexity of precursors will be equivalent for large classes of $J_i$ (corresponding to the Hamiltonian) and $F_i$ (corresponding to the choice of $W_0$). From the above equation, we see that the coordinates, $\beta^k$, are a sum of three terms. The first two terms are only in the $X$ and $Y$ directions. The time dependence is always of the form $Jt$, so any overall rescaling in $J$ can be absorbed into a redefinition of the time parameter $t$. The second term involves the dot product of $\hat{J}_i$ and $F_i$ so it only depends on the angle between $F_i$ and $J_i$. The third term is proportional to the cross product between $\hat J_i$ and $F_i$, so it is in the $Z$ direction and again only depends on the angle between $\hat{J}_i$ and $F_i$. Therefore, since the sub-manifold generated by the $X$ and $Y$ directions is isotropic, up to a rescaling of time, the complexity of the precursor only depends on the angle between $F_i$ and $J_i$.




\begin{thebibliography}{99}

\bibitem{Maldacena} 
  J.~M.~Maldacena,
  ``The Large N limit of superconformal field theories and supergravity,''
  Int.\ J.\ Theor.\ Phys.\  {\bf 38}, 1113 (1999)
  doi:10.1023/A:1026654312961, 10.4310/ATMP.1998.v2.n2.a1
  [hep-th/9711200].
  
\bibitem{Gubser} 
  S.~S.~Gubser, I.~R.~Klebanov and A.~M.~Polyakov,
  ``Gauge theory correlators from noncritical string theory,''
  Phys.\ Lett.\ B {\bf 428}, 105 (1998)
  doi:10.1016/S0370-2693(98)00377-3
  [hep-th/9802109].
  
\bibitem{Witten} 
  E.~Witten,
  ``Anti-de Sitter space and holography,''
  Adv.\ Theor.\ Math.\ Phys.\  {\bf 2}, 253 (1998)
  doi:10.4310/ATMP.1998.v2.n2.a2
  [hep-th/9802150].
  
  
  \bibitem{RT} 
  S.~Ryu and T.~Takayanagi,
  ``Holographic derivation of entanglement entropy from AdS/CFT,''
  Phys.\ Rev.\ Lett.\  {\bf 96}, 181602 (2006)
  doi:10.1103/PhysRevLett.96.181602
  [hep-th/0603001].
  
\bibitem{HRT} 
  V.~E.~Hubeny, M.~Rangamani and T.~Takayanagi,
  ``A Covariant holographic entanglement entropy proposal,''
  JHEP {\bf 0707}, 062 (2007)
  doi:10.1088/1126-6708/2007/07/062
  [arXiv:0705.0016 [hep-th]].
  
  \bibitem{ADH}
A.~Almheiri, X.~Dong and D.~Harlow,
``Bulk Locality and Quantum Error Correction in AdS/CFT,''
JHEP \textbf{04}, 163 (2015)
doi:10.1007/JHEP04(2015)163
[arXiv:1411.7041 [hep-th]].

\bibitem{DHW}
X.~Dong, D.~Harlow and A.~C.~Wall,
``Reconstruction of Bulk Operators within the Entanglement Wedge in Gauge-Gravity Duality,''
Phys. Rev. Lett. \textbf{117}, no.2, 021601 (2016)
doi:10.1103/PhysRevLett.117.021601
[arXiv:1601.05416 [hep-th]].
  
  \bibitem{CompVol} 
  L.~Susskind,
  ``Computational Complexity and Black Hole Horizons,''
  [Fortsch.\ Phys.\  {\bf 64}, 24 (2016)]
  Addendum: Fortsch.\ Phys.\  {\bf 64}, 44 (2016)
  doi:10.1002/prop.201500093, 10.1002/prop.201500092
  [arXiv:1403.5695 [hep-th], arXiv:1402.5674 [hep-th]].

  
  \bibitem{Brown} 
  A.~R.~Brown, D.~A.~Roberts, L.~Susskind, B.~Swingle and Y.~Zhao,
  ``Complexity, action, and black holes,''
  Phys.\ Rev.\ D {\bf 93}, no. 8, 086006 (2016)
  doi:10.1103/PhysRevD.93.086006
  [arXiv:1512.04993 [hep-th]].
  
 \bibitem{Brown2} 
  A.~R.~Brown, D.~A.~Roberts, L.~Susskind, B.~Swingle and Y.~Zhao, 
 ``Holographic Complexity Equals Bulk Action?,''
 Phys.\ Rev.\ Lett. {\bf116}, no. 19, 191301 (2016) 
 [arXiv:1509.07876 [hep-th]].
 

  \bibitem{LPiTP} 
  L.~Susskind,
  ``Three Lectures on Complexity and Black Holes,''
  arXiv:1810.11563 [hep-th].
  
\bibitem{Lehner}
L.~Lehner, R.~C.~Myers, E.~Poisson and R.~D.~Sorkin,
Phys. Rev. D \textbf{94}, no.8, 084046 (2016)
doi:10.1103/PhysRevD.94.084046
[arXiv:1609.00207 [hep-th]].

\bibitem{Hartman}
T.~Hartman and J.~Maldacena,
``Time Evolution of Entanglement Entropy from Black Hole Interiors,''
JHEP \textbf{05}, 014 (2013)
doi:10.1007/JHEP05(2013)014
[arXiv:1303.1080 [hep-th]].

  
\bibitem{Chapman17} 
  S.~Chapman, M.~P.~Heller, H.~Marrochio and F.~Pastawski,
  ``Toward a Definition of Complexity for Quantum Field Theory States,''
  Phys.\ Rev.\ Lett.\  {\bf 120}, no. 12, 121602 (2018)
  doi:10.1103/PhysRevLett.120.121602
  [arXiv:1707.08582 [hep-th]].
  
 \bibitem{Jefferson} 
  R.~Jefferson and R.~C.~Myers,
  ``Circuit complexity in quantum field theory,''
  JHEP {\bf 1710}, 107 (2017)
  doi:10.1007/JHEP10(2017)107
  [arXiv:1707.08570 [hep-th]].
  
\bibitem{Shenker}
S.~H.~Shenker and D.~Stanford,
``Black holes and the butterfly effect,''
JHEP \textbf{03}, 067 (2014)
doi:10.1007/JHEP03(2014)067
[arXiv:1306.0622 [hep-th]].

\bibitem{Stanford}
D.~Stanford and L.~Susskind,
``Complexity and Shock Wave Geometries,''
Phys. Rev. D \textbf{90}, no.12, 126007 (2014)
doi:10.1103/PhysRevD.90.126007
[arXiv:1406.2678 [hep-th]].

\bibitem{Switchback}
L.~Susskind and Y.~Zhao,
``Switchbacks and the Bridge to Nowhere,''
[arXiv:1408.2823 [hep-th]].

\bibitem{QCNC}
A.~R.~Brown, L.~Susskind and Y.~Zhao,
``Quantum Complexity and Negative Curvature,''
Phys. Rev. D \textbf{95}, no.4, 045010 (2017)
doi:10.1103/PhysRevD.95.045010
[arXiv:1608.02612 [hep-th]].

\bibitem{Nielsen1}
M.~A.~Nielsen,
``A geometric approach to quantum circuit lower bounds,''
[arXiv:quant-ph/0502070].

\bibitem{Nielsen2}
M.~A.~Nielsen, M.~Dowling, M.~Gu, and A.~C.~Doherty, 
``Quantum Computation as Geometry,''
Science 311, 1133 (2006), 
[arXiv:quant-ph/0603161].

\bibitem{Nielsen3}
M.~A.~Nielsen, M.~R.~Dowling, M.~Gu, and A.~C.~Doherty, 
``Optimal control, geometry, and quantum computing'', 
Phys. Rev. A 73, 062323 (2006), 
[arXiv:quant-ph/0603160].

\bibitem{Nielsen4}
 M.~R.~Dowling and M.~A.~Nielsen, 
``The geometry of quantum computation,''
[arXiv:quant-ph/0701004].

\bibitem{Nielsen5}
M.~Gu, A.~Doherty, and M.~Nielsen,
``Quantum control via geometry: An explicit example,''
Physical Review A, 78 032327, 
[arXiv:0808.3212 [quant-ph]].

\bibitem{2ndLaw}
A.~R.~Brown and L.~Susskind,
``Second law of quantum complexity,''
Phys. Rev. D \textbf{97}, no.8, 086015 (2018)
doi:10.1103/PhysRevD.97.086015
[arXiv:1701.01107 [hep-th]].

 \bibitem{SingleQubit} 
  A.~R.~Brown and L.~Susskind,
  ``Complexity geometry of a single qubit,''
  Phys.\ Rev.\ D {\bf 100}, no. 4, 046020 (2019)
  doi:10.1103/PhysRevD.100.046020
  [arXiv:1903.12621 [hep-th]].
  
\bibitem{BindComp}
V.~Balasubramanian, M.~DeCross, A.~Kar and O.~Parrikar,
``Binding Complexity and Multiparty Entanglement,''
JHEP \textbf{02}, 069 (2019)
doi:10.1007/JHEP02(2019)069
[arXiv:1811.04085 [hep-th]].
  
\bibitem{SYKComp}
V.~Balasubramanian, M.~Decross, A.~Kar and O.~Parrikar,
``Quantum Complexity of Time Evolution with Chaotic Hamiltonians,''
JHEP \textbf{01}, 134 (2020)
doi:10.1007/JHEP01(2020)134
[arXiv:1905.05765 [hep-th]].




%
%
%





\end{thebibliography}
\end{document}